\documentclass[twocolumn,showpacs,superscriptaddress,prb,amsmath,amssymb,letterpaper]{revtex4}
\usepackage{times}
\usepackage{amsmath,bm,amsfonts}
\usepackage{dcolumn}
\usepackage{graphicx}
\usepackage{latexsym}

\begin{document}

\title{Competing quantum paramagnetic ground states of the Heisenberg antiferromagnet on the star lattice}

\author{Bohm-Jung \surname{Yang}}
\affiliation{Department of Physics, University of Toronto,
Toronto, Ontario M5S 1A7, Canada}

\author{Arun \surname{Paramekanti}}
\affiliation{Department of Physics, University of Toronto,
Toronto, Ontario M5S 1A7, Canada}

\author{Yong Baek \surname{Kim}}
\affiliation{Department of Physics, University of Toronto,
Toronto, Ontario M5S 1A7, Canada}
\affiliation{School of Physics,
Korea Institute for Advanced Study, Seoul 130-722, Korea}

\date{\today}

\begin{abstract}
We investigate various competing paramagnetic ground states of the Heisenberg antiferromagnet on the
two dimensional star lattice which exhibits
geometric frustration. Using slave particle mean field theory combined with a projective symmetry
group analysis, we examine a variety of candidate spin liquid states on this lattice, including chiral spin liquids,
spin liquids with Fermi surfaces of spinons, and nematic spin liquids which break lattice rotational
symmetry. Motivated by connection to large-N SU(N) theory as well as numerical exact diagonalization
studies, we also examine various valence bond solid (VBS) states on this lattice. Based on a study of
energetics using Gutzwiller projected states, we find that a fully gapped spin liquid state is the
lowest energy spin liquid candidate for this model. We also find,
from a study of energetics using
Gutzwiller projected wave functions and bond operator approaches, that this spin liquid is unstable
towards two different VBS states --- a VBS state which respects all the Hamitonian symmetries and a
VBS state which exhibits $\sqrt{3}\times\sqrt{3}$ order ---
depending on the ratio of the Heisenberg
exchange couplings on the two inequivalent bonds of the lattice. We compute the triplon dispersion in both
VBS states within the
bond operator approach and discuss possible implications of our work for future experiments on candidate
materials.
\end{abstract}

\pacs{74.20.Mn, 74.25.Dw}

\maketitle

\section{\label{sec:intro} Introduction}

\begin{figure}[t]
\centering
\includegraphics[width=6.5 cm]{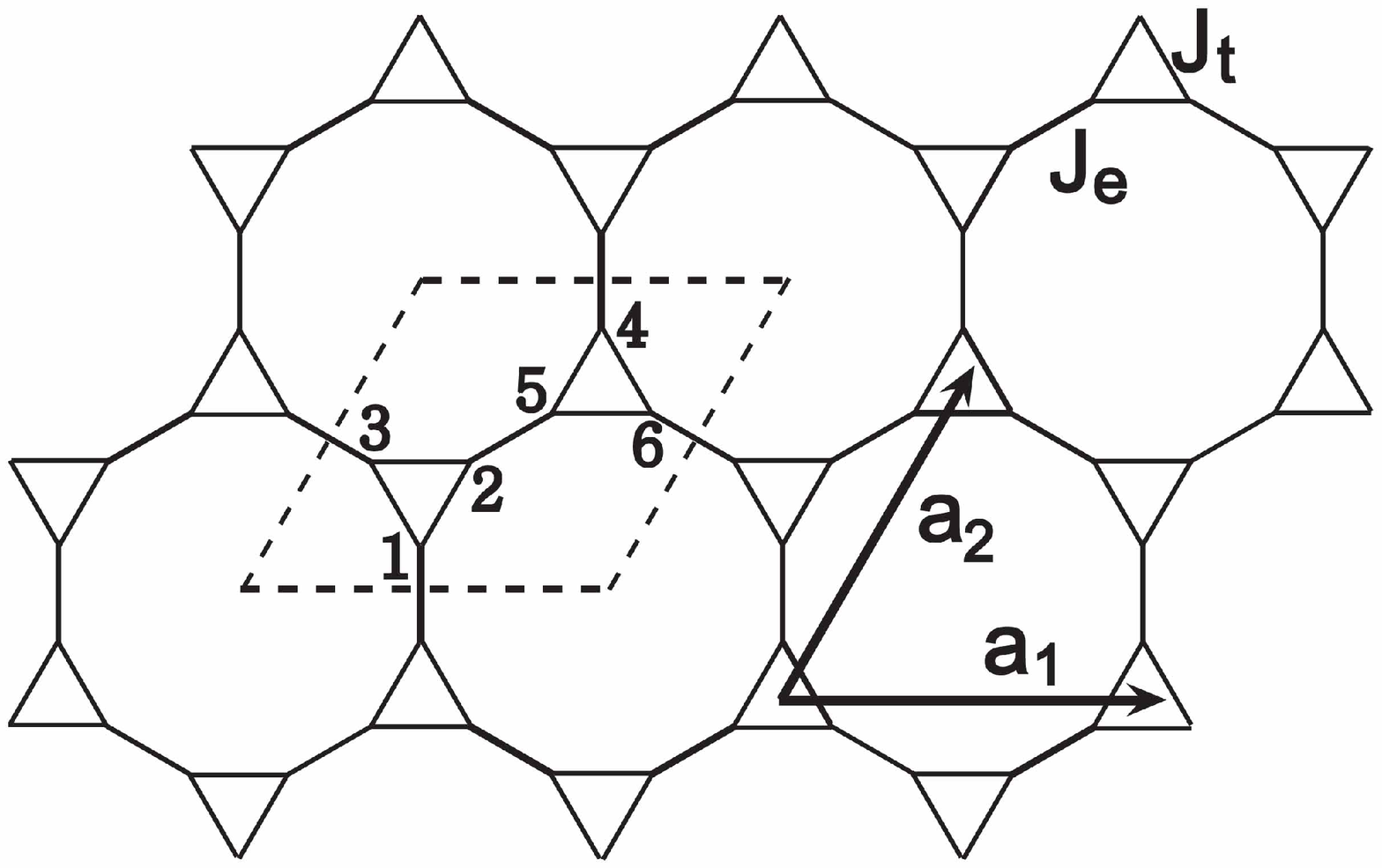}
\includegraphics[width=8 cm]{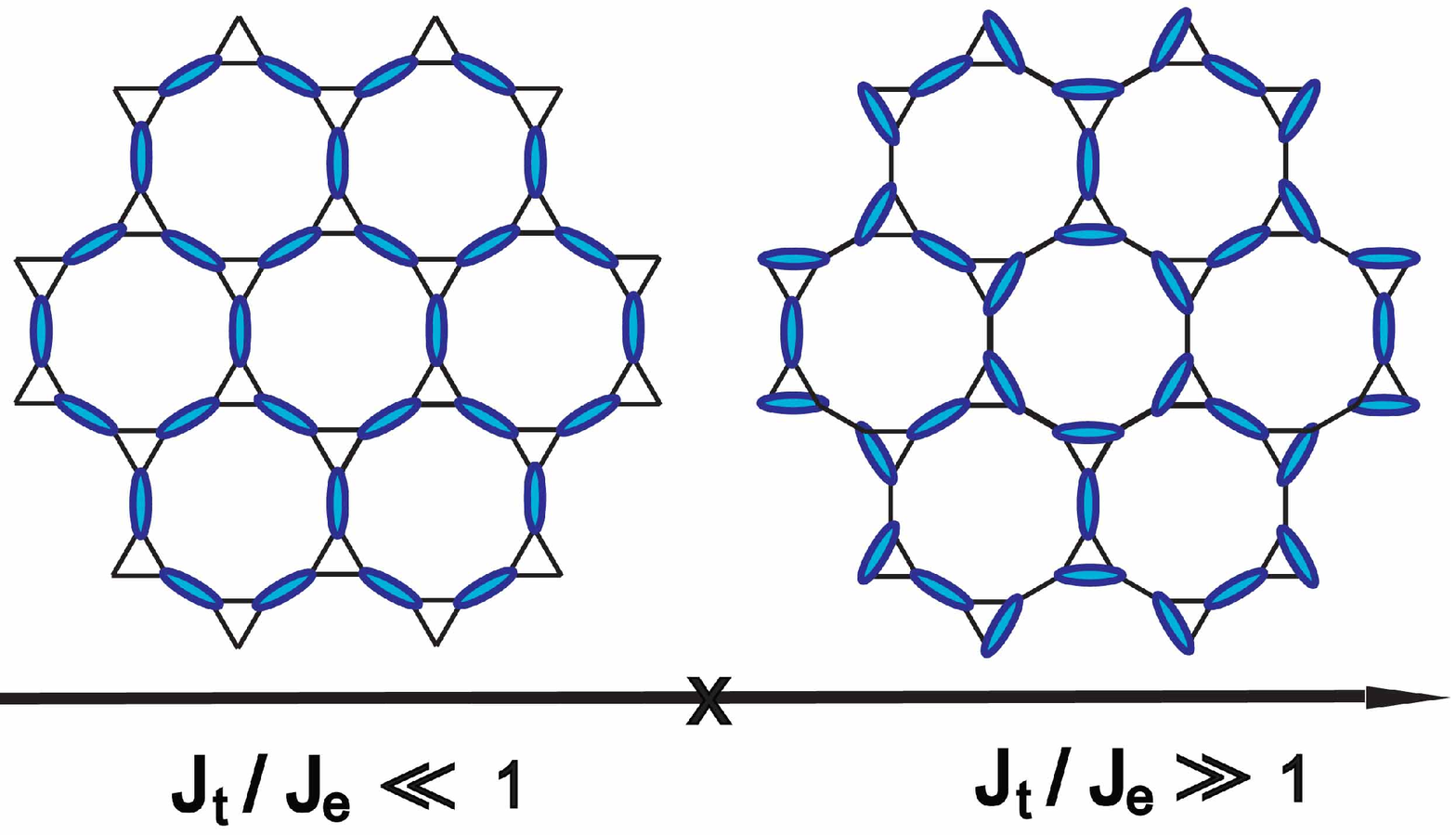}
\caption{(Color online) Top: Structure of the star lattice depicting the six-site unit cell, the chosen lattice basis
vectors $\textbf{a}_{1}$=2$\hat{\textbf{x}}$ and $\textbf{a}_{2}$ = $\hat{\textbf{x}}$ + $\sqrt{3}$$\hat{\textbf{y}}$,
and the bonds with Heisenberg exchange couplings $J_e$ (`expanded bonds') and $J_t$ (`triangle bonds').
Bottom: Phase diagram of the antiferromagnetic Heisenberg model on the star lattice.
For $J_{t}/J_{e}\ll1$, the ground state is a valence bond solid (VBS) phase ($J_{e}$-dimer VBS)
in which every dimer sits on the expanded links connecting neighboring triangles.
For $J_{t}/J_{e}\gg1$, the ground state is a VBS with an 18-site unit cell
(the columnar 18-site VBS).
} \label{fig:LatticeStructure}
\end{figure}

In recent years, several geometrically frustrated spin-1/2 magnets have been synthesized which appear to not order
magnetically even at temperatures well below the characteristic exchange couplings.
Among these are the quasi-two-dimensional (2D) triangular
organic material $\kappa$-BEDT(CN)$_3$, \cite{kanoda} the kagome lattice herbertsmithite, \cite{nocera}
and distorted kagome lattice volborthite,
\cite{hiroi}
and the three-dimensional hyperkagome lattice
magnet \cite{takagi} Na$_4$Ir$_3$O$_8$. A large class of these magnets appear to exhibit gapless spin liquid behavior down to very low
temperatures, leading to the exciting possibility that they may possess exotic ground states with fractionalized
excitations.\cite{lawler,palee,marston,Hastings,Ran,Hermele,Ryu,JHKim,Burnell}
Others among these have been proposed to weakly order into singlet valence bond solid (VBS) states which break
lattice symmetries.\cite{MarstonKagome,Nikolic,Huse,Huse2} As yet, there is no clear picture of what combination of geometric effects and spin
interactions will lead to spin liquid ground states or VBS ground states; this necessitates a theoretical and
experimental exploration of
various new lattice geometries as well as possible ring-exchange
interactions beyond the simplest Heisenberg spin exchange interaction.

In this
paper, we focus on understanding several competing
ground states of the nearest neighbor $S\!=\!1/2$ Heisenberg model
\begin{align}\label{eqn:hamiltonian}
H &= \sum_{\langle i, j \rangle} J_{i,j} \textbf{S}_{i}\cdot \textbf{S}_{j},
\end{align}
on the 2D star lattice, shown in Fig.~1,
as a function of $J_t/J_e$ where $J_t$ and $J_e$ are the exchange couplings on the `triangle bonds' and
`expanded bonds'.
Our motivation for this study is two-fold.
First, the recent synthesis of a `star lattice' organic Iron Acetate quantum magnet \cite{starexpt}
raises the possibility that a $S\!=\!1/2$ variant
may possibly be synthesized in the near future and our results should be applicable to such systems.
Second, this lattice has a sufficiently different geometry from
more commonly studied quantum magnets --- it may be viewed
either as a variant of the kagome lattice or as a decorated honeycomb lattice ---
which allows us to explore the effect of this new lattice geometry on possible
spin liquid physics and valence bond solid phases in quasi-2D systems.

Motivated by interpolating between this spin-1/2 SU(2) model and an SU(N) generalization
at large N which permits a mean field solution with fermionic spinon excitations,\cite{Read,Marston,Marston2}
we examine a large number of interesting U(1) spin liquids as candidate ground states
of the nearest neighbor $S\!=\!1/2$ Heisenberg model on the 2D star lattice. Guided by
earlier work on the kagome lattice,\cite{Ran} we focus
on spin liquid states denoted by SL$[\Phi_\triangle,\Phi_\nabla,\Phi_{dodecagon}]$ where
$\Phi_\triangle$,$\Phi_\nabla$, and $\Phi_{dodecagon}$ denote, respectively, the ``fictitious" fluxes
seen by the fermionic spinons as they move around an elementary plaquette of the
lattice: an up triangular plaquette $\triangle$, a down triangular plaquette $\nabla$, or the $12$-site dodecagon
plaquette. In terms of the original spin variable, the fluxes on the triangular plaquettes
correspond to scalar spin chiralities of the form ${\bf S}_1\cdot {\bf S}_2 \times {\bf S}_3$,
while $\Phi_{dodecagon}$ is related to an operator defined by the twelve
spins around the dodecagon loop.
Depending on the flux values, these
spin liquids represent gapped chiral spin liquids which break time-reversal symmetry,\cite{Wen}
or states with
gapless Fermi surfaces of spinons, or gapped spin liquids with no
broken symmetries.

From a study of energetics of various flux values using Gutzwiller projected wave function numerics
for the physical case of $N=2$, we
show that a particular gapped spin liquid, which we denote as SL$[0,0,\pi]$, which does not break
lattice or time-reversal symmetries emerges as a favorable candidate over a wide range of $J_t/J_e$.
This is in striking contrast to earlier work on the kagome lattice from two perspectives.
First, as we show, the effect of projection is far more dramatic on the star lattice
when compared with the kagome lattice; a numerical Gutzwiller projection of the mean field states leads to
a complete reordering of the energies of the candidate spin liquids.
Second, unlike the kagome lattice
case where the lowest energy variational state of this form is a spin liquid with massless Dirac
fermion excitations,\cite{Ran}
the SL$[0,0,\pi]$  is a gapped U(1) spin liquid --- we therefore know that it is ultimately unstable towards spinon
confinement at low energies\cite{Polyakov} unlike the Dirac fermion states whose stability depends on the number of
fermion flavors $N$.\cite{HermeleASL}

We find that the SL$[0,0,\pi]$ state
naturally forms strong dimers on the `expanded
bonds' for small values of $J_t/J_e$ thus leading to a confined state, a $J_e$-dimer VBS,
which respects all symmetries
of the Hamiltonian.\cite{Richter} For large $J_t/J_e$, numerical
exact diagonalization (ED) studies of this model carried out in a restricted nearest-neighbor dimer basis
showed signatures of $\sqrt{3}\times\sqrt{3}$ ordering.\cite{Misguich}
We argue that another motivation to study possible dimer orders that break lattice symmetries
is that such ordering often appear quite naturally
in the large-N fermionic SU(N) theory
as recognized in the early work of Affleck and Marston\cite{Marston,MarstonKagome} and
shown in various other models studied
recently.\cite{Georges,MarstonAP} Inspired by these results, we consider
various candidate VBS phases from different perspectives --- a large-N route, a bond operator formalism, and
Gutzwiller projected wave function numerics. All of these point to a transition to a $\sqrt{3}\times\sqrt{3}$
ordered VBS phase for large enough $J_t/J_e \gtrsim 2-2.5$, leading us to the phase diagram shown in
Fig.~1. Our result is in broad agreement with the ED study although
the transition point estimated from our work is somewhat larger than the exact diagonalization (ED)
result which yields $(J_t/J_e)_{\rm crit} \approx 1.3$; the ED result may, however, may suffer from significant
finite size effects. We then discuss possible routes by which the SL$[0,0,\pi]$ state might be unstable towards
such $\sqrt{3}\times\sqrt{3}$ VBS order instead of the $J_e$-dimer VBS.
We present results for the triplon dispersion in both VBS states which can be tested in
inelastic neutron scattering experiments on candidate materials.

Finally, although the various other interesting
spin liquids we study do not appear to be energetically viable ground states for
the nearest neighbor Heisenberg model on the star lattice, they have energies which are close to the
ground state. They might thus be stabilized as ground states by small changes in the Hamiltonian, such
as further neighbor exchange or spin-phonon coupling,
or they might be relevant to understanding the intermediate
energy scale properties
or finite temperature physics of
materials which realize this model. We therefore elucidate some of the properties of these spin liquid states.

This paper is organized as follows. We begin, in Section \ref{sec:meanfield}
by formulating the mean field theory of the Heisenberg model
on this lattice in a slave particle description using fermionic spinons which we relate to a large-N SU(N)
approach. Based on this, we classify and study the physical properties of a number of candidate spin
liquid ansatzes. We next turn, in Section \ref{sec:dimer}, to a study of dimerized states on this lattice,
and present a group theoretic classification of $\sqrt{3}\times\sqrt{3}$ orders as well as a large-N
justification of specific candidate VBS phases. Section \ref{sec:energetics} contains a discussion of the energetics
of various spin liquid states using mean field theory as well as a Gutzwiller projected wave function study
and bond operator approaches of candidate VBS phases. Section V discusses the various ways in which the
SL$[0,0,\pi]$ state, which is the lowest energy spin liquid state, might be unstable towards VBS ordering
as a result of spinon interactions and from a Gutzwiller wave function approach.
The phase transition between VBS phases is described in Section \ref{sec:phasetransition}.
Section \ref{sec:triplon}
contains a discussion of the triplon dispersion in both VBS states which we think are realized in this
model. We conclude with a discussion about experimental implications
in Section \ref{sec:discussion}. Details of various calculations are contained in Appendices A-C.

\section{\label{sec:meanfield} Spin liquid phases on the star lattice}

\subsection{\label{sec:formulation} Formulation of the mean field theory}

We investigate the ground state of the S=1/2 Heisenberg antiferromagnet on the
star lattice, which can be described by the following Hamiltonian,
\begin{align}\label{eqn:hamiltonian}
H &= \sum_{\langle i, j \rangle} J_{i,j} \textbf{S}_{i}\cdot \textbf{S}_{j},
\end{align}
where $i$ and $j$ indicate the positions of nearest neighbor spin pairs.
Natural description of this lattice system requires the consideration of two
inequivalent links, that is, one link lying on a triangle (a triangular link) and the other link
connecting two neighboring triangles (an expanded link).
We assign two different exchange couplings $J_{e}$ and $J_{t}$ on expanded
and triangular links, respectively.
We can also label a site $i$ by pairs ($\textbf{R}$, $n$) where $\textbf{R}$ denotes the
location of a unit cell and $n$ labels the six sites inside a single unit cell.
(See Fig.~\ref{fig:LatticeStructure}.)

To construct spin liquid states we introduce the fermionic spinon operators,
$f_{\sigma}$ ($\sigma = \uparrow or \downarrow$) to represent the spin operator;

\begin{align}
S^{\alpha}_{i} &= \frac{1}{2} \sum_{\sigma_{1}, \sigma_{2}} f^{\dag}_{i, \sigma_{1}}\sigma^{\alpha}_{\sigma_{1}, \sigma_{2}} f_{i, \sigma_{2}},
\quad (\alpha=x,y,z).
\end{align}

Since this representation alone contains unphysical local configurations enlarging the Hilbert space, we have to impose the following
 local constraint, $f^{\dag}_{\uparrow}f_{\uparrow} + f^{\dag}_{\downarrow}f_{\downarrow}$ = 1,
to recover the physical Hilbert space.
Using the fermioninc spinon representation of the spin operator, the Heisenberg spin
Hamiltonian can be rewritten as follows.

\begin{align}
H &= - \sum_{\sigma_{1}, \sigma_{2}}\sum_{<ij>} \frac{J_{ij}}{2} f^{\dag}_{i, \sigma_{1}} f_{j, \sigma_{1}} f^{\dag}_{j, \sigma_{2}} f_{i, \sigma_{2}}.
\end{align}
Here we have dropped unimportant constant terms.

To decouple the four fermion interaction term we define spin singlet order parameters,
$\chi_{ij}$ $\equiv$ $\frac{1}{2}$$\sum_{\sigma}$ $\langle f^{\dag}_{i, \sigma}f_{j, \sigma} \rangle$.
After imposing the single occupancy constraint using the Lagrange multipliers $\mu_{i}$,
the mean field Hamiltonian is given by

\begin{align}
H_{MF} = - & \sum_{\sigma}\sum_{<ij>} J_{ij}( f^{\dag}_{i, \sigma} f_{j, \sigma} \chi^{*}_{ij} + h.c. )
+ \sum_{<ij>}2J_{ij}|\chi_{ij}|^{2} \nonumber \\
\qquad &+ \sum_{i, \sigma} \mu_{i} ( f^{\dag}_{i, \sigma} f_{i, \sigma} - 1 ).
\end{align}

To describe the phase fluctuation of the mean field ansatz, we express $\chi_{ij}$ as
$\chi_{ij}$ = $\overline{\chi}_{ij}$ $e^{i a_{ij}}$, which leads to
the following Hamiltonian,
\begin{align}
H_{U(1)} = - & \sum_{\sigma}\sum_{<ij>} J_{ij}( f^{\dag}_{i, \sigma} f_{j, \sigma} \overline{\chi}_{ij}e^{-i a_{ij}} + h.c. )
\nonumber \\
\qquad &+ \sum_{i, \sigma} \mu_{i} ( f^{\dag}_{i, \sigma} f_{i, \sigma} - 1 ).
\end{align}

In the above Hamiltonian $H_{U(1)}$, the local U(1) gauge symmetry of the spin Hamiltonian
which comes from the local conservation of the fermion number
is manifest via the following gauge transformation.\cite{SLbook,SLreview}

\begin{align}
f_{i} & \rightarrow f_{i} e^{i \theta_{i}},\nonumber \\
a_{ij} & \rightarrow a_{ij} - \theta_{i} + \theta_{j}.
\end{align}
Here $a_{ij}$ describing the phase fluctuation of $\chi_{ij}$
plays the role of the spatial components of the U(1) gauge field.
Namely, we have reformulated the quantum spin model
as the problem of the spinons strongly coupled to the U(1) gauge field.

A systematic way of studying the coupled spinon and gauge field system is
to consider the large-N reformulation of the problem extending
the spin SU(2) symmetry to SU(N) (with N even).\cite{Marston, Marston2}
We let the flavor index $\alpha$ run from 1 to N and modify the
single occupancy constraint as,
\begin{align}
\sum_{\alpha}f^{\dag}_{i,\alpha}f_{i,\alpha} = \frac{N}{2}.
\end{align}

In addition, we scale the interaction strength $J_{ij}$/2 to be  $J_{ij}$/N to make
each term of the Hamiltonian to be of order N. The resulting large-N Hamiltonian
is given by

\begin{align}\label{eqn:largeNHamiltonian}
H &= - \sum_{\alpha, \beta = 1}^{N}\sum_{<ij>} \frac{J_{ij}}{N} f^{\dag}_{i, \alpha} f_{j, \alpha} f^{\dag}_{j, \beta} f_{i, \beta}.
\end{align}

To treat the quartic interactions we perform a mean field decoupling
by introducing SU(N) singlet valence bond ,
$\chi_{ij}$ $\equiv$ $\frac{1}{N}$$\sum_{\alpha}$ $\langle f^{\dag}_{i, \alpha}f_{j, \alpha}\rangle$.
Assuming the valence bond amplitude is a complex number,
we obtain the mean field Hamiltonian given by

\begin{align}
H_{MF} = - & \sum_{\alpha}\sum_{<ij>} J_{ij}( f^{\dag}_{i, \alpha} f_{j, \alpha} \chi^{*}_{ij} + h.c. )
+ N \sum_{<ij>} J_{ij}|\chi_{ij}|^{2} \nonumber \\
\qquad &+ \sum_{i, \alpha} \mu_{i} ( f^{\dag}_{i, \alpha} f_{i, \alpha} - \frac{N}{2} ).
\end{align}


Since the fluctuations of $\chi_{ij}$
and the average local density  $\frac{1}{N}$$\sum_{\alpha}$ $\langle f^{\dag}_{i, \alpha}f_{i, \alpha}\rangle$
scale as 1/$\sqrt{N}$, we can safely neglect those fluctuations
in the large-N limit justifying the mean field approximation.


Here we consider the following mean field ansatz
$\overline{\chi}_{ij}$ = $|\chi_{ij}|$ $e^{i \phi_{ij}}$ where
$|\chi_{ij}|$ = $\chi_{e}$ on expanded links
and $|\chi_{ij}|$ = $\chi_{t}$ on triangular links, respectively.
We specify the various flux patterns inside the elementary
plaquettes, i.e., the triangles and the dodecagons.
The flux inside a triangle $\Phi_{\triangle}$, for example, is defined in the
following way, $e^{i \Phi_{\triangle}}$ $\equiv$ $e^{i (\phi_{ij} + \phi_{jk} + \phi_{ki})}$
where $\langle ijk \rangle$ indicates the three corners of a triangle
traversed along the counterclockwise direction. The flux inside
a dodecagon is also defined in the same manner.
Since the flux inside a closed loop is a gauge invariant object,
different spin liquid ansatz can be distinguished based on the
flux values inside the elementary plaquettes. In particular,
we use the term SL$[\Phi_{\triangle}, \Phi_{\nabla},\Phi_{dodecagon}]$
to represent the ansatz which has the fluxes $\Phi_{\triangle}$ inside an up-pointing triangle,
$\Phi_{\nabla}$ inside a down-pointing triangle and $\Phi_{dodecagon}$ inside a dodecagon.
With a given flux configuration we determine $|\chi_{ij}|$ and $\mu_i$
self-consistently by solving the following coupled mean field equations,

\begin{align}
&\frac{1}{N_{site}}\sum_{i}\sum_{\alpha}\langle f^{\dag}_{i,\alpha}f_{i,\alpha}\rangle = \frac{N}{2},\nonumber\\
&\chi_{ij} = \frac{1}{N} \sum_{\alpha} \langle f^{\dag}_{i, \alpha}f_{j, \alpha}\rangle,
\end{align}
where $N_{site}$ is the number of lattice sites.

\subsection{\label{sec:properties} Properties of competing spin liquid phases}
In this section we discuss
the characteristic properties of various spin liquid phases and their instabilities.
In particular, we focus on translationally invariant mean field states
which have nonzero $|\chi_{ij}|$ on every link of the lattice.
Extensive discussion on possible dimerized phases is given later in Sec.~\ref{sec:dimer}.
As shown in the previous studies about the spin liquid phases on the square\cite{Ran2} and kagome\cite{marston, Fa} lattices,
the inclusion of additional spin interactions can change the relative energetics of
different spin liquid phases.
Therefore it is useful to understand the nature of various competing
spin liquid states which are the potential ground states
of spin Hamiltonians beyond the nearest neighbor Heisenberg model.
\\

(a) \underline{SL$[0,0,0]$ : Uniform spin liquid state}
\\

\begin{figure}[t]
\centering
\includegraphics[width=8 cm]{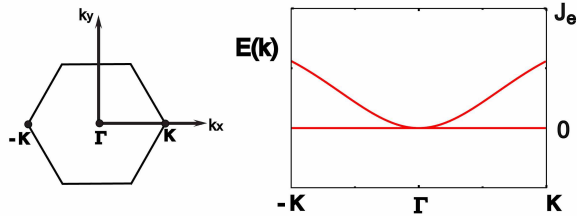}
\caption{(Color online)
The spinon dispersion of the SL[0,0,0] near the Fermi level.
Among the six bands inside the first Brillouin zone
we plot the third and four band which are lying close to the Fermi energy
along the $k_{x}$ axis.
The flat band is lying at the Fermi level which touches with the dispersive
band at the $\Gamma$ point.
} \label{fig:band000}
\end{figure}

To describe this state we introduce two real mean field order parameters,
$\chi_{e}$ and $\chi_{t}$, which lie on the expanded and
triangular links, respectively. Since the unit cell contains
six sites we obtain six different bands inside the first Brillouin zone.
Among the six bands, the third and fourth bands near
the Fermi level show an interesting structure displayed in Fig.~\ref{fig:band000}.
There is a flat band lying on the Fermi energy which is touching another
dispersive band at the zone center, $\Gamma(\textbf{k}=\textbf{0})$.
The flat band comes from
the existence of the localized eigenstates, which occur due to the destructive
interference of hopping amplitudes between the localized units.\cite{Bergman}
The flat band of the uniform spin liquid
on the kagome lattice emerges owing to the same reason.
However, in contrast to the kagome lattice problem,
the flat band is lying exactly at the Fermi level on the
star lattice.

The flatness of the band at the Fermi energy is not the generic property
of the uniform spin liquid. There are perturbations
which do not break any lattice symmetry but spoil the flatness by generating
curvature. The third nearest neighbor hopping is such an example.
However, the quadratic degeneracy at the zone center is protected by
the point group symmetry of the underlying unit cell.

To understand the stability of the spin liquid we derive the low energy
effective Hamiltonian, which describes the states near the zone center,
expanding the Hamiltonian up to the quadratic order of the momentum $\textbf{k}$.
The procedure for deriving the effective Hamiltonian is outlined in the Appendix~\ref{sec:eff000}.
The resulting Hamiltonian is given by,
\begin{align}
H_{\text{eff}} = &\frac{1}{m_{\text{eff}}}\int \frac{d^{2}\textbf{k}}{(2\pi)^2} \psi^{\dag}(\textbf{k})
h_{\text{eff}}(\textbf{k})\psi(\textbf{k}),\nonumber
\end{align}
in which
\begin{align}
h_{\text{eff}}(\textbf{k}) = &
(k^{2}_{x}+k^{2}_{y})\tau_{0}-(k^{2}_{x}-k^{2}_{y})\tau_{z}-2k_{x}k_{y}\tau_{x},\nonumber
\end{align}
where the Pauli matrix $\tau_{i}$ is acting on the two-component
space of the continuum field $(\psi)^{T}$=$(\psi_{1},\psi_{2})$
which describes the two low energy states near the $\Gamma$ point.

The SL[0,0,0] state respects all the space group symmetry of
the lattice. In particular, if we choose the gauge in which
$\chi_{ij}$=$\chi_{t}$ on every triangular link and  $\chi_{ij}$=$\chi_{e}$
on every expanded link with $\chi_{t}$ and $\chi_{e}$ being real constants,
the action of the symmetry generators on the spin operator, $S_{i}$,
is the same as that on the spinon operator, $f_{i,\sigma}$.
Since we consider the low energy excitations near the zone center, we focus on the
action of the point group symmetry on the continuum fields.
The $D_{6}$ point group of the star lattice consists of twelve
symmetry generators and is generated by the two elements, $C_{\frac{\pi}{3}}$ and $R_{y}$.
Here $C_{\frac{\pi}{3}}$ means the $\frac{\pi}{3}$ rotation with respect to the
center of a dodecagon and $R_{y}$ indicates the reflection about
the x-axis. The details on the elements of the $D_{6}$ point group
are discussed in the Sec.~\ref{sec:grouptheory}.

Under the $C_{\frac{\pi}{3}}$ and $R_{y}$, the continuum field $\psi$ transforms
in the following way,

\begin{align}\label{eq:symmetryof000}
C_{\frac{\pi}{3}} : \psi & \longrightarrow e^{-\imath \frac{\pi}{3}\tau_{y}} \psi, \nonumber \\
R_{y} : \psi & \longrightarrow  \tau_{z} \psi,
\end{align}
meaning all the fermion bilinears $\psi^{\dag}\tau_{\alpha}\psi$ ($\alpha$ = x, y, z)
are forbidden by the point group symmetry.
Note that the $\psi^{\dag}\tau_{y}\psi$ breaks time-reversal symmetry as well
since it shows sign change under complex conjugation.

Next we consider the fermion bilinear terms that contain spatial derivatives.
Because the dynamical critical exponent is two, the terms with a single
spatial derivative are relevant and those with two spatial derivatives
are marginal perturbations.
Investigating the transformation rule under the $D_{6}$ point group
symmetry, it can be easily checked that $\psi^{\dag}\tau_{y}\psi$ transforms
as the one dimensional $A_{2}$ irreducible representation and
($\psi^{\dag}\tau_{x}\psi$,$\psi^{\dag}\tau_{z}\psi$) forms a basis
for the two dimensional $E_{2}$ irreducible representation.\cite{Tinkham}
Similarly, the transformation properties of derivative terms
can be determined.
At first, the linear derivative term, ($\partial_{x}$,$\partial_{y}$)
transforms as a two dimensional $E_{1}$ irreducible representation.
To have terms with linear derivatives in the Hamiltonian,
the product of the fermion bilinear and the derivative
must be invariant under the point group symmetry operations.
Using the decompositions of $E_{1}\bigotimes A_{2}=E_{1}$
and  $E_{1}\bigotimes E_{2}=B_{1}\bigoplus B_{2}\bigoplus E_{1}$,
we see that every product of fermion bilinears and the linear derivative
is not invariant under the point group symmetry.
Therefore linear derivative terms are not allowed.
On the other hand, we have a second derivative term
($\partial_{x}^{2}-\partial_{y}^{2}$,$\partial_{x}\partial_{y}$)
making a two dimensional $E_{2}$ irreducible
representation. Using
$E_{2}\bigotimes A_{2}=E_{2}$
and  $E_{2}\bigotimes E_{2}=A_{1}\bigoplus A_{2}\bigoplus E_{2}$,
we see that there is a term following $A_{1}$ irreducible representation, which is nothing but
$\psi^{\dag}[(2\partial_{x}\partial_{y})\tau_{x}+(\partial_{x}^{2}-\partial_{y}^{2})\tau_{z}]\psi$.
Therefore in addition to the isotropic $\psi^{\dag}(\partial_{x}^{2}+\partial_{y}^{2})\psi$ term,
$\psi^{\dag}[(2\partial_{x}\partial_{y})\tau_{x}+(\partial_{x}^{2}-\partial_{y}^{2})\tau_{z}]\psi$
is the only term allowed by symmetry. Since these terms are already present
in the Hamiltonian, the low energy properties of the SL[0,0,0]
are not spoiled by these marginal perturbations. However, these
perturbations add curvature to the flat band.

Finally, we discuss the effect of the four fermion interaction
terms on the stability of the SL[0,0,0] state. Though a simple power
counting shows that the four fermion interactions are marginal,
they are actually marginally relevant. Recently,
the effects of the four fermion interaction
on the quadratic band crossing are studied using the renormalization
group approach.\cite{kaisun2,Vafek}
According to Ref.~\onlinecite{kaisun2}, the leading weak coupling
instability leads to the state with nonzero $\langle\psi^{\dag}\tau_{y}\psi\rangle$, breaking the time reversal symmetry.
It means that SL[0,0,0] state is unstable toward a chiral spin liquid
state supporting chiral edge states.
\\

(b) \underline{SL$[0,0,\pi]$ state}
\\

\begin{figure}[t]
\centering
\includegraphics[width=6.5 cm]{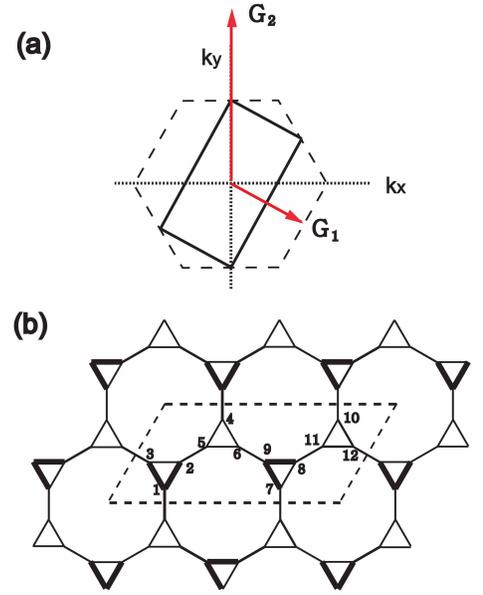}
\caption{(Color online)
(a) The reduced Brillouin zone (the solid line) corresponding
to the doubled unit cell along the $\textbf{a}_{1}$ direction.
$\textbf{G}_{1}$ and $\textbf{G}_{2}$ indicate the reciprocal lattice vector
corresponding to the doubled unit cell.
(b) The flux configuration of the SL$[0,0,\pi]$ state.
The twelve-site unit cell is surrounded by a dotted box.
We assign -1 (+1) for the hopping amplitude on the thick (thin) bond.
} \label{fig:gauge00pi}
\end{figure}

\begin{figure}[t]
\centering
\includegraphics[width=8.5 cm]{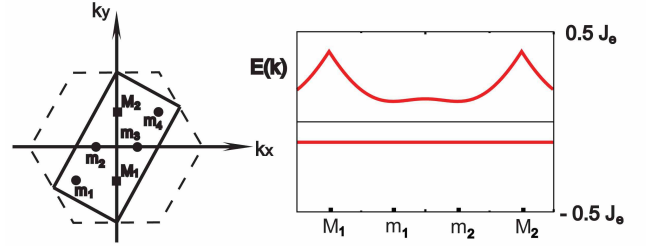}
\caption{(Color online)
The low energy spinon bands of the SL$[0,0,\pi]$ state
near the Fermi level.
The lower band (valence band) is flat and doubly degenerate.
The upper band (conduction band) has small dispersion. The locations
of the conduction band minimum (maximum) are described by $\textbf{m}_{i}$ ($\textbf{M}_{i}$).
The dispersion is plotted along the $k_{y}$=$\sqrt{3}$$k_{x}$+$\frac{\pi}{2\sqrt{3}}$
line passing the $\textbf{m}_{1}$ and $\textbf{m}_{2}$.
} \label{fig:band00pi}
\end{figure}

The SL[0,0,$\pi$] state supports $\pi$ flux piercing the dodecagons. Even though
this state does not break the translational symmetry, the mean field
description requires doubling of the unit cell. Here we consider
the doubling of the unit cell along the $\textbf{a}_{1}$ direction.
For the lattice vectors 2$\textbf{a}_{1}$ and $\textbf{a}_{2}$,
the reciprocal lattice vectors are given by
\begin{align}
\textbf{G}_{1} =& (\frac{\pi}{2}, -\frac{\pi}{2\sqrt{3}}),\quad
\textbf{G}_{2} = (0, \frac{2 \pi}{\sqrt{3}}).\nonumber
\end{align}
The reduced Brillouin zone corresponding to the above reciprocal
lattice vectors is depicted in Fig.~\ref{fig:gauge00pi}(a).

For the mean field description of the SL$[0,0,\pi]$ state,
we have chosen the flux configuration as described in Fig.~\ref{fig:gauge00pi}(b).
Since there are twelve sites inside the unit cell, we have
twelve bands within the Brillouin zone.
The mean field spinon dispersion of the low energy bands near the Fermi level
is described in Fig.~\ref{fig:band00pi}.
This state does not have a spinon Fermi surface and shows a gapped
spectrum. The lower flat band (valence band) is doubly degenerate
and the upper band (conduction band) is dispersive.

According to the projected wave function study that is discussed in detail later
in Sec.~\ref{sec:VMC},
the SL[0,0,$\pi$]
state has the lowest ground state energy among the various
spin liquid ansatz over a wide parameter range.
Unfortunately, however, the SL[0,0,$\pi$] state is unstable
once gauge fluctuation is allowed.
Since the spinon spectrum has a finite gap, the low energy excitations
are described by the compact U(1) gauge theory. In 2 + 1 dimension,
the compact U(1) gauge theory is confining,\cite{Polyakov} which means that free spinons
with unit gauge charge can only make charge neutral bound states.
In addition, the interaction between spinons can also induce
various kinds of broken symmetry states.
Extensive discussion on the instability of the SL[0,0,$\pi$] state
and its relation with candidate valence bond solid phases are given in Sec.~\ref{sec:sl00pi}.
\\

(c) \underline{SL$[\frac{\pi}{2} ,\frac{\pi}{2} , \pi]$ : A chiral spin liquid state}
\\

\begin{figure}[t]
\centering
\includegraphics[width=8 cm]{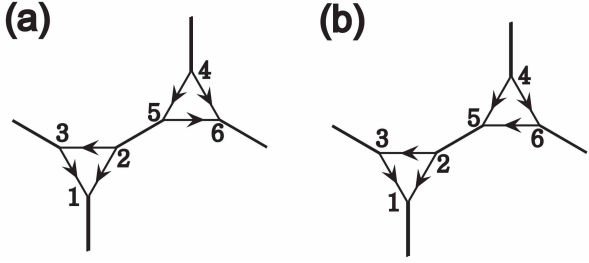}
\caption{(Color online)
Flux configuration of the ansatz SL$[\frac{\pi}{2} ,\frac{\pi}{2} , \pi]$
and  SL$[-\frac{\pi}{2} ,\frac{\pi}{2} , 0]$. Hopping along (against) the direction
of an arrow induce a phase $\pi$/2 (-$\pi$/2).
(a) The flux pattern for SL$[\frac{\pi}{2} ,\frac{\pi}{2} , \pi]$.
Counterclockwise motion along an triangular link results in the flux $\frac{\pi}{2}$
(b) The flux pattern for SL$[-\frac{\pi}{2} ,\frac{\pi}{2} , 0]$.
Counterclockwise motion along an up-pointing (down-pointing) triangular link
results in the flux -$\frac{\pi}{2}$ ($\frac{\pi}{2}$).
} \label{fig:fluxconfig}
\end{figure}
Next we consider flux phases which possess finite flux inside
triangles.  A triangle that supports $\frac{\pi}{2}$ flux
breaks time-reversal and parity symmetry but preserves the combination
of them. A convenient way to investigate the time-reversal symmetry breaking in
spin systems is to consider the expectation value of the scalar
spin chirality operator\cite{Wen} defined as follows,

\begin{align}
\hat{C}_{ijk}& \equiv \textbf{S}_{i} \cdot ( \textbf{S}_{j} \times \textbf{S}_{k} ).
\end{align}
Since $\hat{C}_{ijk}$ is odd under both the time reversal ($T$) and parity operations,
the ground state breaks both symmeties when the expectation
value of the scalar spin chirality operator  $\langle \hat{C}_{ijk} \rangle$ is nonzero.
In other words, the scalar spin chirality plays the role of the
order parameter measuring time reversal symmetry breaking.

Because the unit cell contains two triangles, we can define the following two
different scalar spin chirality operators,

\begin{align}
\hat{C}_{uniform}& \equiv \textbf{S}_{1} \cdot ( \textbf{S}_{2} \times \textbf{S}_{3} )
+\textbf{S}_{4} \cdot ( \textbf{S}_{5} \times \textbf{S}_{6} ), \nonumber \\
\hat{C}_{staggered}& \equiv \textbf{S}_{1} \cdot ( \textbf{S}_{2} \times \textbf{S}_{3} )
-\textbf{S}_{4} \cdot ( \textbf{S}_{5} \times \textbf{S}_{6} ),
\end{align}
where $\hat{C}_{uniform}$ and $\hat{C}_{staggered}$ are the uniform and staggered
scalar spin chiralities, respectively.
To understand the symmetry properties of the $\hat{C}_{uniform}$ and $\hat{C}_{staggered}$
we have to recognize that there are two different reflection symmetries on the star lattice.
The reflection ($P_{1}$) with respect to the axis connecting the center of a dodecagon
with the mid-point of an expanded link (for example, the $a$ axis in Fig.~\ref{fig:twofold}(a))
interchanges the up-pointing triangles and the down-pointing triangles.
On the other hand, the other reflection ($P_{2}$) about the axis connecting
the center of a dodecagon with a vertex of a triangle, (for example, the $A$ axis in Fig.~\ref{fig:twofold}(a)),
leaves each triangle as it is. Under the $P_{2}$ reflection both
$\hat{C}_{uniform}$ and $\hat{C}_{staggered}$ change their signs. Here
we use the term parity to indicate the $P_{1}$ reflection symmetry. Note that the parity
operation is equivalent to the reflection in two dimensional space.

$\hat{C}_{uniform}$ is odd under both
the time-reversal ($T$) and parity ($P_{1}$) but even under inversion ($I$).
On the other hand $\hat{C}_{staggered}$ is odd under time-reversal and
inversion but even under parity. Here the inversion operation
is defined with respect to the mid-point of the expanded link connecting two
triangles.
In both cases, the combination of the time-reversal, parity transformation,
and inversion ($T \cdot P_{1} \cdot I$) is equivalent to the identity operation under which
both the $\hat{C}_{uniform}$ and $\hat{C}_{staggered}$ are invariant.\cite{kaisun1}

\begin{figure}[t]
\centering
\includegraphics[width=8 cm]{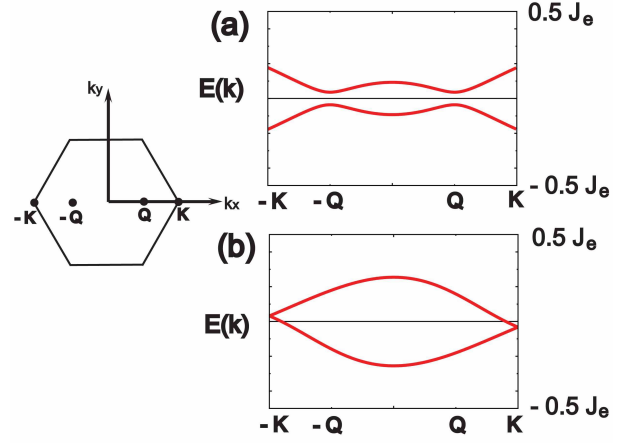}
\caption{(Color online)
The mean field spinon band structure of the ansatz SL$[\frac{\pi}{2} ,\frac{\pi}{2} , \pi]$
and  SL$[-\frac{\pi}{2} ,\frac{\pi}{2} , 0]$ along the $k_{x}$ axis when $J_{t}$=2$J_{e}$.
(a) The first Brillouin zone. Here Q and -Q denote the points in which
minimum (maximum) of the conduction (valence) band  of the SL$[\frac{\pi}{2} ,\frac{\pi}{2} , \pi]$ occurs.
The K and - K indicate the points where the linear band touching between
the two bands near the Fermi energy of the SL$[-\frac{\pi}{2} ,\frac{\pi}{2} , 0]$ occurs.
(b) The dispersion of the valence and conduction band corresponding to the
SL$[\frac{\pi}{2} ,\frac{\pi}{2} , \pi]$ ansatz.
The low energy excitation near the $\pm$Q can be described
by the massive Dirac particles.
(c) The dispersions of the two bands near the Fermi level for SL$[-\frac{\pi}{2} ,\frac{\pi}{2} , 0]$.
The low energy excitation near the $\pm$K point can be described by the Dirac particles
under the staggered chemical potential.
} \label{fig:fluxband}
\end{figure}

The SL$[\frac{\pi}{2} ,\frac{\pi}{2} , \pi]$ is characterized by nonzero
$\hat{C}_{uniform}$ but with vanishing $\hat{C}_{staggered}$. Therefore
it breaks time-reversal and parity transformation while it is invariant
under the combined operation.  It is a chiral spin liquid state which has a
finite energy gap. In Fig.~\ref{fig:fluxband}(a) we plot the spinon dispersion
near the Fermi level
corresponding to the valence and conduction bands. The energy gap is minimum
at the momentum $\textbf{Q}$=($\pi$/3,0) and -$\textbf{Q}$.
If we expand the mean field Hamiltonian near the dispersion
minimum, $\textbf{Q}$ and -$\textbf{Q}$
in the limit of large $J_{t} / J_{e}$, we can get the following
effective low energy Hamiltonian,

\begin{align}\label{eq:EqDirac1}
H_{eff} =& \int \frac{d^{2}q}{(2\pi)^2} \Psi^{\dag}(q) \big\{v_{F}[q_{x}\tau_{x}+q_{y}\tau_{y}]-m\tau_{z}\big\}\Psi(q),
\end{align}
where the Fermi velocity $v_{F}$ = $J_{e}\chi_{e}/\sqrt{3}$ and the mass
$m$ = $(J_{e}^{2}\chi_{e}^{2}) / (\sqrt{3}J_{t}\chi_{t})$.
In the above we define the eight component Dirac fermion field,
$\Psi^{\dag} = (\varphi^{\dag}_{1, \alpha, \sigma},\varphi^{\dag}_{2, \alpha, \sigma})$
in which 1 and 2 are the two-component Dirac indices, $\alpha$ and $\sigma$
are indices for the nodes ($\pm \textbf{Q}$) and spins. The Pauli matrix $\tau_{\nu}$
acts on the two-component Dirac space. For later convenience we define
two additional Pauli matrices, $\vec{\mu}$ and $\vec{\sigma}$ acting on
the nodal and spin spaces, respectively.

Since the mass term has the
same sign in the two nodal positions,
integrating out fermions leads to the Chern-Simons gauge field action.
As a consequence, the charge neutral spinon Hall conductivity
should be finite.
The Chern-Simons term stabilizes
the spin liquid ground state by providing a finite mass to the U(1) gauge field. Therefore
the U(1) gauge field can only mediate
a short range interaction between the spinons, which makes
the fractionalized particles (the spinons) to be the elementary
excitations of the spin liquid ground state.\cite{SLbook}
\\

(d) \underline{SL$[-\frac{\pi}{2} ,\frac{\pi}{2} , 0]$ : A nematic spin liquid state}
\\

The SL$[-\frac{\pi}{2} ,\frac{\pi}{2} , 0]$ is characterized by nonzero
$\hat{C}_{staggered}$ but with vanishing $\hat{C}_{uniform}$. Therefore it breaks
both the time-reversal and inversion operation but is invariant under
the parity transformation. Because the fluxes of the two triangles
within the unit cell have opposite sign, the six-fold
rotational symmetry is broken down to the three-fold symmetry.
(See Fig.~\ref{fig:fluxconfig}(b).)
Thus it is a nematic spin liquid.

The mean field spinon dispersion corresponding to the two bands near the
Fermi energy is plotted in Fig.~\ref{fig:fluxband} (b).
The spin liquid ansatz has a spinon Fermi
surface which consists of an electron pocket at the $\textbf{K}$ = ($2\pi / 3$,0) point and
a hole pocket at the - $\textbf{K}$ point.

Expanding the mean field Hamiltonian using $J_{e}\chi_{e} / J_{t}\chi_{t}$
as an expansion parameter, the following effective low energy Hamiltonian
can be obtained,

\begin{align}\label{eq:effectiveH2}
H_{eff} =& \int \frac{d^{2}q}{(2\pi)^2} \Psi^{\dag}(q) \big\{v_{F}[q_{x}\tau_{x}+q_{y}\tau_{y}]-M\mu_{z}\big\}\Psi(q),
\end{align}
where the fermi velocity $v_{F}$ = $J_{e}\chi_{e}/\sqrt{3}$ and the ``staggered" field
$M$ = $(J_{e}^{2}\chi_{e}^{2}) / (\sqrt{3}J_{t}\chi_{t})$.
Since the effective chemical potentials coming from the ``staggered" field $M$
have the opposite signs at the two nodal points, we have both an electron
pocket (at the $\textbf{K}$ point) and a hole pocket (at the -$\textbf{K}$ point)
on the fermi surface.

In contrast to the SL$[\frac{\pi}{2} ,\frac{\pi}{2} , \pi]$ state which
has a gapped spinon spectrum, the SL$[-\frac{\pi}{2} ,\frac{\pi}{2} , 0]$ state
has gapless low energy excitations. To confirm that the low energy
description based the above effective Hamiltonian in Eq.~(\ref{eq:effectiveH2})
is valid after including the fluctuation beyond the mean field description,
we have to check whether there are relevant perturbations which are allowed
by symmetry. Especially, some of the fermion bilinears, which are made of $\Psi$,
can potentially generate various mass terms which spoil the low energy
description of Eq.~(\ref{eq:effectiveH2}).

To judge the stability of this spin liquid state,
we have to understand how the symmetries of the microscopic Hamiltonian
are realized in the effective continuum theory. Even though the original
spin Hamiltonian is invariant under the full space group transformations,
after the gauge theory formulation of the problem, the symmetry
of the mean field Hamiltonian is realized projectively. That is,
under the symmetry transformation $S$ with the mapping $\textbf{i}$$\rightarrow$$S(\textbf{i})$,
the spinon operator $f_{\textbf{i}, \sigma}$ transforms in the following way,
\begin{align}
S : & f_{\textbf{i}, \sigma} \rightarrow G_{S}(\textbf{i})f_{S(\textbf{i}), \sigma}, \nonumber
\end{align}
where $G_{S}(\textbf{i})$ is a phase factor which depends on the symmetry operation  $S$,
and a local coordinate $\textbf{i}$. The group of the symmetry operations
which make the mean field Hamiltonian invariant is called the projective
symmetry group (PSG).\cite{Wen_PSG,Fa_PSG}

\begin{figure}[t]
\centering
\includegraphics[width=5 cm]{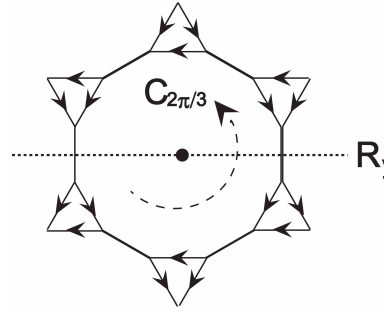}
\caption{(Color online)
The generators of the point group symmetry of the SL$[-\frac{\pi}{2} ,\frac{\pi}{2} , 0]$.
$R_{y}$ maps y to - y while $C_{2 \pi / 3}$ induce the rotation by $2 \pi /3$
with respect to the center of the dodecagon.
}\label{fig:plusminussymmetry}
\end{figure}

To perform the PSG analysis we have to specify the symmetry group
of the spin Hamiltonian. The star lattice has the $D_{6}$ point group
symmetry generated by the six-fold rotation symmetry with respect to
the center of a dodecagon and the reflections.
However, due to the finite fluxes inside triangles,
the SL$[-\frac{\pi}{2} ,\frac{\pi}{2} , 0]$ state breaks some
$D_{6}$ point group symmetries.
Especially, the six-fold rotational symmetry is broken down to three-fold rotational symmetry.
The point group symmetry of the SL$[-\frac{\pi}{2} ,\frac{\pi}{2} , 0]$ state
is generated by the 2$\pi$/3 rotation ($C_{2 \pi /3}$) around
the center of the dodecagon and the reflection ($R_{y}$) which maps $y$ to - $y$.
The symmetry operations which generate the point group of the SL$[-\frac{\pi}{2} ,\frac{\pi}{2} , 0]$ state
are depicted in Fig.~\ref{fig:plusminussymmetry}.
The SL$[-\frac{\pi}{2} ,\frac{\pi}{2} , 0]$ state is also invariant under
the translations ($T_{a1}$ and $T_{a2}$) by the lattice vectors $\textbf{a}_{1}$ and $\textbf{a}_{2}$.
In combination with the above point group symmetry, the translational symmetry defines the space group
of the  SL$[-\frac{\pi}{2} ,\frac{\pi}{2} , 0]$ state.
In addition,
the SL$[-\frac{\pi}{2} ,\frac{\pi}{2} , 0]$ state is invariant under the combination ($T\cdot I$) of the time-reversal ($T$)
and inversion ($I$) as well as the spin rotation. Finally, it has the
charge conjugation symmetry ($C^{*}$) via the mapping $f_{i \alpha}$ $\rightarrow$
$\epsilon_{i}f^{\dag}_{i \alpha}$ where $\epsilon_{i}$ = 1
for i=1,2,3 and -1 for i=4,5,6. Under these symmetry operations
the continuum field $\Psi$ transforms as follows.

\begin{align}\label{eq:symmetryofcontinuum}
T \cdot I : \Psi & \longrightarrow (\imath \sigma_{y}) \tau_{x} \Psi, \nonumber \\
C^{*} : \Psi & \longrightarrow (\imath \tau_{x}) [\Psi^{\dag}]^{T}, \nonumber \\
T_{a_{1}} : \Psi & \longrightarrow e^{-\imath \frac{2 \pi}{3}\mu_{z}} \Psi, \nonumber \\
T_{a_{2}} : \Psi & \longrightarrow e^{\imath \frac{2 \pi}{3}\mu_{z}} \Psi, \nonumber \\
R_{y} : \Psi & \longrightarrow  \mu_{z} \tau_{x} \Psi, \nonumber \\
C_{\frac{2 \pi}{3}} : \Psi & \longrightarrow e^{-\imath \frac{2 \pi}{3}\tau_{z}} \Psi.
\end{align}

Using the above transformation rules we can easily check that
$\Psi^{\dag} \mu_{z} \Psi$ is the only fermion bilinear
which is allowed by symmetry. Therefore the low energy Hamiltonian
in Eq.~(\ref{eq:effectiveH2}) is valid even after we include
the fluctuations and protected by the projective symmetry group.
Some details about how we have determined the transformation rule
of the continuum fields are explained in the Appendix~\ref{sec:fieldsymmetry}.

Next we discuss about possible instability
of the SL$[-\pi/2,\pi/2,0]$ state. The low energy effective
Hamiltonian (Eq.~(\ref{eq:effectiveH2})) which is obtained from the
perturbative expansion in powers of $J_{e}\chi_{e} / J_{t}\chi_{t}$
implies that the electron pocket (at the $\textbf{K}$ point)
and the hole pocket (at the $-\textbf{K}$ point) are nested in the large $J_{t}$/$J_{e}$ limit.
Therefore the instability in the particle-hole channel with the momentum $2\textbf{K}$
is expected.
The following two fermion bilinears,
$\hat{m}_{x}$=$\Psi^{\dag}\tau_{z}\mu_{x}\Psi$ and $\hat{m}_{y}$=$\Psi^{\dag}\tau_{z}\mu_{y}\Psi$,
are especially important in this respect.
Addition of the mass term $H_{M}=M_{x}\hat{m}_{x}+M_{y}\hat{m}_{y}$ to the effective Hamiltonian in Eq.~(\ref{eq:effectiveH2})
leads to the mass gap of $2 \sqrt{M^{2}+M_{x}^{2}+M_{y}^{2}}$.
Since these mass terms are anticommuting with the effective Hamiltonian in Eq.~(\ref{eq:effectiveH2}),
the pair ($\hat{m}_{x}$,$\hat{m}_{y}$) opens the largest mass gap than any
other pairs of possible mass terms.
Interestingly, ($\hat{m}_{x}$,$\hat{m}_{y}$) transforms
nontrivially under the space group operations.
Its symmetry property is consistent with some ordered state
with $\sqrt{3}\times \sqrt{3}$-type translational symmetry breaking.
Using the terminology defined in Sec.~\ref{sec:grouptheory},
($\hat{m}_{x}$,$\hat{m}_{y}$) transforms as the $E_{3}$ irreducible representation
under the
enlarged point group $G_{P,\textbf{b}}$.
The detailed discussion on the group theory for the star lattice
is given in Sec.~\ref{sec:grouptheory}.
\\

(e) \underline{SL$[\frac{\pi}{2},\frac{\pi}{2},0]$ and SL$[-\frac{\pi}{2},\frac{\pi}{2},\pi]$}
\\

\begin{figure}[t]
\centering
\includegraphics[width=8.5 cm]{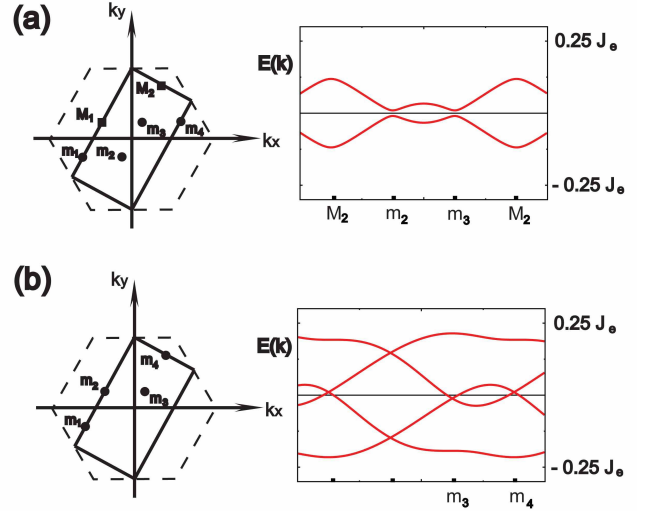}
\caption{(Color online)
The low energy spinon bands of the  SL$[\frac{\pi}{2},\frac{\pi}{2},0]$
and SL$[-\frac{\pi}{2},\frac{\pi}{2},\pi]$ plotted along the $k_{y}$=$\sqrt{3}$$k_{x}$
direction.
(a) For SL$[\frac{\pi}{2},\frac{\pi}{2},0]$.
The locations of the minimum (maximum) band gap are described by $m_{i}$ ($M_{i}$).
(b) For SL$[-\frac{\pi}{2},\frac{\pi}{2},\pi]$.
Here $m_{1}$ and $m_{3}$ ($m_{2}$ and $m_{4}$)
indicate the location of the electron (hole) pockets.
} \label{fig:bandhalfpi}
\end{figure}

We obtain SL$[\frac{\pi}{2},\frac{\pi}{2},0]$ (SL$[-\frac{\pi}{2},\frac{\pi}{2},\pi]$) phase by putting
additional $\pi$ flux on every dodecagon on top of
SL$[\frac{\pi}{2},\frac{\pi}{2},\pi]$ (SL$[-\frac{\pi}{2},\frac{\pi}{2},0]$) states.
Due to the introduction of the additional $\pi$ flux,
the mean field description requires unit cell doubling
although the actual physical wave function maintains the translational invariance.

The spinon dispersion of the SL$[\frac{\pi}{2},\frac{\pi}{2},0]$ state
is described in Fig.~\ref{fig:bandhalfpi}(a).
Basically, the structure of the low energy spectrum
of SL$[\frac{\pi}{2},\frac{\pi}{2},0]$ is similar to that of SL$[\frac{\pi}{2},\frac{\pi}{2},\pi]$,
except that the number of the momentum points which support low energy excitations is doubled.
Both of them are characterized by finite $\hat{C}_{uniform}$ indicating
the time-reversal and parity symmetry breaking.
Therefore the SL$[\frac{\pi}{2},\frac{\pi}{2},0]$ state is
also a chiral spin liquid state.
The low energy excitations can be
described by the effective Hamiltonian similar to Eq.~(\ref{eq:EqDirac1})
which can be obtained following the same procedure
we used to derive Eq.~(\ref{eq:EqDirac1}) for SL$[\frac{\pi}{2},\frac{\pi}{2},\pi]$.

In Fig.~\ref{fig:bandhalfpi}(b) we have drawn the low energy spinon
excitation spectrum of the SL$[-\frac{\pi}{2},\frac{\pi}{2},\pi]$ state.
There are two electron pockets (around $\textbf{m}_{1}$ and $\textbf{m}_{3}$)
and two hole pockets (around $\textbf{m}_{2}$ and $\textbf{m}_{4}$).
It is characterized by finite $\hat{C}_{staggered}$ showing
broken time reversal and inversion symmetry. Since the fluxes
inside up-triangles and down-triangles have opposite signs,
the six-fold rotational symmetry is broken down to three-fold
rotational symmetry. Therefore it is another nematic
spin liquid state.


\section{\label{sec:dimer}  Dimer phases}

\subsection{\label{sec:largeN} Large-N approach}

According to the pioneering work by D.S.Rokhsar,\cite{Rokhsar} when the lattice system
is dimerizable, the best mean field
ansatz is one of dimerized states in the large-N limit of the SU(N)-generalized
Heisenberg model.
Here we call a lattice to be dimerizable when it is possible to make
every site belong to a dimer
and a lattice site
be paired with one and only one of its neighboring site.
In particular, when every dimer is lying on the link which has the maximum
exchange coupling $J_{max}$, the dimer state belongs to the ground state manifold of the
mean field Hamiltonian.
In terms of the variable $\chi_{ij}$, we have finite
$\overline{\chi}_{ij}$ only on the dimers lying on the link which has
the maximum spin coupling $J_{max}$.

\begin{figure}[t]
\centering
\includegraphics[width=4 cm]{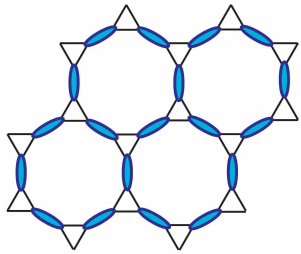}
\caption{(Color online) (a) The $J_{e}$-dimer VBS state. This is
the ground state in the
large-N limit when $J_{e}$ $>$ $J_{t}$.
} \label{fig:Jedimer}
\end{figure}

The star lattice is dimerizable with respect to $J_{e}$.
Therefore when $J_{e}$ is larger than $J_{t}$, it has a unique
dimerized ground state (we call it the $J_{e}$-dimer VBS) in which every dimer is lying on an
expanded link connecting neighboring triangles.
In Fig.~\ref{fig:Jedimer} we describe the geometric arrangement of singlet dimers
of the $J_{e}$-dimer VBS phase.

On the other hand, the Rokhsar's general theorem cannot be applied
when $J_{t}$ is larger than $J_{e}$. This is because the star lattice
is not dimerizable with respect to the $J_{t}$ links and every dimer configuration
defined on the star lattice contains a finite number of dimers lying on the $J_{e}$ links.
Therefore it is possible that the translationally invariant mean field ansatz can be the ground state
even in the large-N limit.

\begin{figure}[t]
\centering
\includegraphics[width=6 cm]{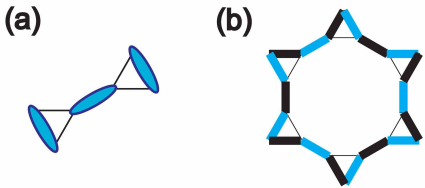}
\caption{(Color online)
(a) The representative local dimer configuration when $J_{t}$ $>$ $J_{e}$.
A dimer lying on an expanded link connects two neighboring filled triangles
which are supporting dimers on them.
(b) The 18-site flippable loop which consists of the alternating bright(blue)
and dark(black) thick lines.
} \label{fig:Jtdimer}
\end{figure}

When $J_{t}$ $>$ $J_{e}$, we have to maximize the number of the dimers
lying on triangular links to minimize the ground state energy of dimerized states.
Since every triangle can support a single dimer at most (we call the triangle with a dimer lying on it
a filled triangle), the remaining unpaired lattice
point of the filled triangle has to be a part of the dimer lying on an expanded link.
In other wards, every dimer lying on an expanded link is connecting two filled triangles
and this describes a representative local dimer configuration of the lowest energy
dimerized states when $J_{t}$ $>$ $J_{e}$. (See Fig.~\ref{fig:Jtdimer}(a).)
Using this local dimer configuration
as a building block we can construct infinite number of degenerate dimerized ground
states.

To understand how the degeneracy of these dimerized phases is lifted by fluctuations,
we consider the 1/N corrections to the ground state energy.
In Ref.~\onlinecite{Read}, Read and Sachdev investigated the 1/N correction
systematically for a similar problem defined on the square lattice. We first review the main ideas
of their work briefly and extend the approach to our star lattice problem.
For a given dimer configuration, $\bar{\chi}_{ij}$ we include
the fluctuations $\delta \chi_{ij}$ as
$\chi_{ij}$ = $\bar{\chi}_{ij}$ + $\delta \chi_{ij}$.
Here $\bar{\chi}_{ij}$ is nonzero only on the link supporting a dimer lying on it.
Expanding
the effective action to the quadratic order in the fluctuations,
the ground state degeneracy of the dimerized states on the square lattice
could be lifted by the following
terms, $\delta S_{eff}$ $\propto$ N $\bar{\chi}_{ij}$$\delta\chi_{jk}$$\bar{\chi}_{kl}$$\delta\chi_{li}$.
Here $i,j,k,$ and $l$ indicate the four corners of a square plaquette.
When a pair of links lying in parallel are occupied by two dimers
($\bar{\chi}_{ij}$ and $\bar{\chi}_{kl}$) and the remaining pair of the links
are assigned to the fluctuations ($\delta\chi_{jk}$ and $\delta\chi_{li}$),
the $\delta S_{eff}$ term above can induce additional lowering of the ground state energy.
It means that dimer configurations which support the maximum
number of the parallel dimer pairs span the ground state manifold.
The four-fold degenerate columnar valence bond solid is selected as the ground
state following these procedures.

The above idea of the 1/N-correction can also be rephrased
in the following way. For every square plaquette composed of
two parallel dimers, we can define a loop which consists
of alternating occupied and empty links.
Here we call such a loop as a flippable loop\cite{Nikolic}
because two degenerate dimer configurations are connected via
a loop flip, i.e., the interchange of the occupied and empty links.
The 1/N-correction captures the energy lowering through the resonance
process which can also be described as a loop flip.
The resulting ground state (a columnar dimer state) supports
the maximum number of the flippable loops.
This idea can be generalized to the higher order corrections
and the degeneracy of dimerized states begins to be lifted
from the lowest order correction corresponding to the smallest
flippable loop.
Marston and Zeng\cite{Marston} discussed the effect of the 1/N-correction
on the degeneracy lifting process for the kagome lattice antiferromagnet.
There, the first term that lifts the degeneracy
involves the six-site flippable loop, which is the so-called
perfect hexagon with three dimers on it. The valence bond
solid ground state of the kagome lattice, which contains 36-site
within the unit cell, results from the condition of maximizing the number
of the perfect hexagons.\cite{Marston, Nikolic, Huse}
The similar idea was also applied to the square-kagome antiferromagnet.\cite{Georges}

In the star lattice problem with $J_{t}$ $>$ $J_{e}$, the dimerized ground
states are constructed by repeating the representative local dimer configuration
displayed in Fig.~\ref{fig:Jtdimer}(a). In this ground state manifold,
the smallest flippable loop contains 18 sites with a dodecagon at the center,
which is shown in Fig.~\ref{fig:Jtdimer}(b). Here when the bright (blue) thick
link is occupied by a dimer, the neighboring dark (black) thick link is empty
and vice versa. By interchanging the roles played by the bright (blue) links
and the dark (black) links, two degenerate dimerized phases can be connected.

\begin{figure}[t]
\centering
\includegraphics[width=8 cm]{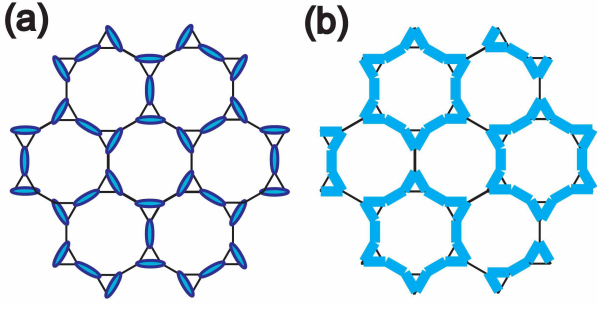}
\caption{(Color online)
Two low energy valence bond solid (VBS) order which are coming from
the 1/N correction.
(a) The columnar 18-site VBS.
(b) The box 18-site VBS.
} \label{fig:Jtdimerconfig}
\end{figure}

Therefore the fluctuation corrections pick the patterns that
maximize the number of the 18-site dimer units participating in
the 18-site flippable loops. In Fig.~\ref{fig:Jtdimerconfig}(a)
we show the valence bond solid order which has the maximum number
of the 18-site flippable loops. Among the six neighboring dodecagons
around an 18-site dimer unit, three can be the centers of the 18-site dimer units.
This is in contrast to the case of the kagome lattice problem.
There, none of the six neighboring hexagons around a perfect hexagon can be
perfect hexagons. In fact, the valence bond solid states on the star
lattice have similarity with those on the square lattice.
In the case of the square lattice, among
the four neighboring square plaquettes around a central plaquette
supporting two parallel dimers, half of them (two square plaquettes)
can support two parallel dimers. By maximizing the number of
the square plaquette composed of two parallel dimers, the columnar
valence bond solid emerges.

Based on the similarity with the square lattice problem, we can call
the valence bond solid in  Fig.~\ref{fig:Jtdimerconfig}(a) as
a columnar 18-site valence bond solid. In this figure all the
dodecagons except the central one support the 18-site dimer unit.
The three-fold degeneracy of the columnar 18-site VBS comes from
the broken translation symmetry.
In addition we also consider another low energy valence bond order which
is displayed in Fig.~\ref{fig:Jtdimerconfig}(b). Here the 18 links
around the 18-site unit have the same finite value of $\bar{\chi}_{ij}$.
In analogy with the square lattice problem this phase can be called
as the 18-site box VBS phase. The translational symmetry breaking
results in the three-fold degeneracy in this phase as well. In the recent numerical
study by G. Misguich et al.\cite{Misguich}, this phase was suggested as a
possible valence bond solid ground state when $J_{t}$ $>$ 1.3 $J_{e}$.
We expect that the 1/N corrections would select the 18-site columnar VBS
as the ground state that can maximize the resonance energy gain from
the smallest flippable loop. However, since the length of the flippable loop
is quite large compared to those on the square
and kagome lattice problem, the energy difference of the two candidate valence bond solids
in Fig.~\ref{fig:Jtdimerconfig} could be very small.

\subsection{\label{sec:grouptheory} Group theoretical approach to $\sqrt{3}\times\sqrt{3}$ bond orders}
The columnar and box 18-site VBS phases discussed in the above section
break the lattice translational symmetry and are described by the enlarged
$\sqrt{3}\times\sqrt{3}$ unit cell.
Here we perform the detailed symmetry analysis
on the bond order that are compatible with the
$\sqrt{3}\times\sqrt{3}$ enlarged unit cell.
\\


\begin{figure}[t]
\centering
\includegraphics[width=8 cm]{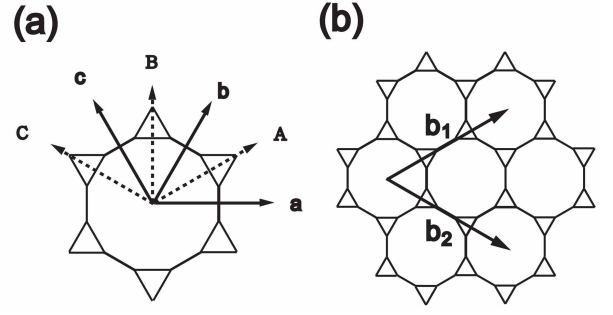}
\caption{(Color online)
(a) The intersections between the reflection planes and the lattice plane.
(b) The lattice vectors $\textbf{b}_{1}$ and $\textbf{b}_{2}$
corresponding to the $\sqrt{3}\times\sqrt{3}$ ordered states.
} \label{fig:twofold}
\end{figure}

(a) \underline{Group theory for the star lattice}
\\
\\
The star lattice has the $D_{6}$ point group symmetry. The twelve elements
of the $D_{6}$ group are as follows.
\begin{align}
D_{6} =& \{E, C_{6}, C_{6}^{2}, C_{6}^{3}, C_{6}^{4}, C_{6}^{5}, R_{a},R_{b},R_{c},R_{A},R_{B},R_{C} \}, \nonumber
\end{align}
where $C_{6}$ means the rotation by 2$\pi$/6 around the center of a dodecagon
and $E$ is the identity operator.
$R_{\alpha}$ indicates the reflection with respect to a plane
orthogonal to the lattice plane. The reflection planes (denoted by $\alpha$) are described in the Fig.~\ref{fig:twofold}(a).

The space group, $G_{S}$, of the star lattice is generated by the translation
group, $G_{T}$,
and the $D_{6}$ point group. An element of the space group can
be written using the Seitz operator $\{g_{D_{6}}|\textbf{t}\}$, where $g_{D_{6}}$
is an element of the $D_{6}$ group and $\textbf{t}$=$n_{1}\textbf{a}_{1}$+$n_{2}\textbf{a}_{2}$.
($n_{1}$ and $n_{2}$ are integers.) The action of a Seitz operator
on a lattice point $\textbf{r}$ is defined as
$\{g_{D_{6}}|\textbf{t}\}\textbf{r}=g_{D_{6}}\textbf{r}+\textbf{t}$.
Note that the translation group, $G_{T}$, is an invariant subgroup
of the space group $G_{S}$ and the point group $D_{6}$ is
the corresponding factor group, i.e., $D_{6}$= $G_{S}$/$G_{T}$.

To understand the symmetry of the enlarged unit cell, we define another
translation group, $G_{T,\textbf{b}}$, whose elements can be written as,
\begin{align}
G_{T,\textbf{b}} =& \{\{E|n_{1}\textbf{b}_{1}+n_{2}\textbf{b}_{2}\};n_{1},n_{2} \in \mathbb{Z} \}
\end{align}
where $\textbf{b}_{1}$ = $\textbf{a}_{1}+\textbf{a}_{2}$ and $\textbf{b}_{2}$ = $2 \textbf{a}_{1}-\textbf{a}_{2}$
are the lattice vectors corresponding to the $\sqrt{3}\times\sqrt{3}$ ordered state.(See Fig.~\ref{fig:twofold}(b))
Since $G_{T,\textbf{b}}$ is an invariant subgroup of the space group $G_{S}$,
the enlarged point group $G_{P,\textbf{b}}$ can be defined as the factor
group $G_{P,\textbf{b}}$= $G_{S}$/$G_{T,\textbf{b}}$. Therefore the elements
of the space group can be written as $\{ g_{G_{P,\textbf{b}}}$ $|$ $n_{1} \textbf{b}_{1}+n_{2}\textbf{b}_{2}\}$
in which $g_{G_{P,\textbf{b}}}$ is an element of the enlarged point group, $G_{P,\textbf{b}}$.
(Note that similar approach was used by Hermele et al., \cite{Hermele} to investigate the symmetry properties
of the object invariant under the translations by 2$\textbf{a}_{1}$ and 2$\textbf{a}_{2}$ on the kagome lattice.)

The construction of the $G_{P,\textbf{b}}$ group is straightforward, whose
elements can be written using the Seitz operator $\{g_{D_{6}}|\bar{\textbf{t}}\}$ with
$\bar{\textbf{t}}$=0, $\textbf{a}_{1}$, $\textbf{a}_{2}$. The 36 elements of the
$\{g_{D_{6}}|\bar{\textbf{t}}\}$ can be grouped into the nine conjugate classes,

\begin{align}
C_{E}=&\{\{ E | 0 \}\}, \nonumber \\
C_{T}=&\{\{ E | \textbf{a}_{1} \},\{ E | \textbf{a}_{2} \} \}, \nonumber \\
C_{6}^{2}=&\{\{ C_{6}^{2} | 0 \},\{  C_{6}^{4} | 0 \} \}, \nonumber \\
C_{6T}=&\{\{ C_{6} | 0 \},\{  C_{6}^{5} | 0 \}, \{ C_{6} | \textbf{a}_{1}\},\{  C_{6}^{5} |  \textbf{a}_{1} \},
\{ C_{6} | \textbf{a}_{2}\},\{  C_{6}^{5} |  \textbf{a}_{2} \} \}, \nonumber \\
C_{6T}^{2}=&\{\{ C_{6}^{2} | \textbf{a}_{1}\},\{  C_{6}^{4} |  \textbf{a}_{1} \},
\{ C_{6}^{2} | \textbf{a}_{2}\},\{  C_{6}^{4} |  \textbf{a}_{2} \} \}, \nonumber \\
C_{6T}^{3}=&\{\{ C_{6}^{3} | 0 \}, \{ C_{6}^{3} | \textbf{a}_{1}\},\{  C_{6}^{3} |  \textbf{a}_{2} \} \}, \nonumber \\
R_{1}=&\{\{ R_{a} | 0 \},\{ R_{b} | 0 \},\{ R_{c} | 0 \} \}, \nonumber \\
R_{1T}=&\{\{ R_{a} | \textbf{a}_{1} \},\{ R_{b} | \textbf{a}_{1} \},\{ R_{c} | \textbf{a}_{1} \},
\{ R_{a} | \textbf{a}_{2} \},\{ R_{b} | \textbf{a}_{2} \},\{ R_{c} | \textbf{a}_{2} \} \}, \nonumber \\
R_{2T}=&\{\{ R_{A} | 0 \},\{ R_{B} | 0 \},\{ R_{C} | 0 \},
\{ R_{A} | \textbf{a}_{1} \},\{ R_{B} | \textbf{a}_{1} \},\{ R_{C} | \textbf{a}_{1} \},\nonumber \\
&\{ R_{A} | \textbf{a}_{2} \},\{ R_{B} | \textbf{a}_{2} \},\{ R_{C} | \textbf{a}_{2} \} \}. \nonumber
\end{align}

\begin{table}
\begin{tabular}{@{}|c|c|c|c|c|c|c|c|c|c|}
\hline \hline
& $C_{E}$ & $C_{T}$ & $C_{6}^{2}$ & $C_{6T}$ & $C_{6T}^{2}$  & $C_{6T}^{3}$ & $R_{1}$ & $R_{1T}$ & $R_{2T}$ \\
\hline \hline
$A_{1}$ & 1 & 1 & 1 & 1 & 1 & 1 & 1 & 1 & 1 \\
$A_{2}$ & 1 & 1 & 1 & 1 & 1 & 1 & -1 & -1 & -1 \\
$B_{1}$ & 1 & 1 & 1 & -1 & 1 & -1 & 1 & 1 & -1 \\
$B_{2}$ & 1 & 1 & 1 & -1 & 1 & -1 & -1 & -1 & 1 \\
$E_{1}$ & 2 & 2 & -1 & 1 & -1 & -2 & 0 & 0 & 0 \\
$E_{2}$ & 2 & 2 & -1 & -1 & -1 & 2 & 0 & 0 & 0 \\
$E_{3}$ & 2 & -1 & 2 & 0 & -1 & 0 & 2 & -1 & 0 \\
$E_{4}$ & 2 & -1 & 2 & 0 & -1 & 0 & -2 & 1 & 0 \\
$Q$     & 4 & -2 & -2 & 0 & 1 & 0 & 0 & 0 & 0 \\
\hline \hline
\end{tabular}
\caption{The Character table of the enlarged point group $G_{P,\textbf{b}}$}
\label{table:charactertable}
\end{table}

Table~\ref{table:charactertable} displays the character table corresponding
to the enlarged point group $G_{P,\textbf{b}}$. It consists of the four
one dimensional irreducible representations ($A_{1}$, $A_{2}$, $B_{1}$, and $B_{2}$)
and the four two dimensional irreducible representations, $E_{\alpha}$ ($\alpha$=1,2,3, and 4)
and a four dimensional representation $Q$.
Note that the four one-dimensional irreducible representations and
the two dimensional representations $E_{1}$ and $E_{2}$ are simple
extensions of the six irreducible representations of the original $D_{6}$
point group. They describe states which are invariant under the
lattice translations by $\textbf{a}_{1}$ or $\textbf{a}_{2}$ as
is reflected in the column for $C_{T}$ in Table.~\ref{table:charactertable}.
Therefore the  $\sqrt{3}\times\sqrt{3}$ type orderings can be
described only through the remaining three irreducible representations
$E_{3}$, $E_{4}$ and $Q$. It turns out that the two dimensional
irreducible representations $E_{3}$ and $E_{4}$ are
especially important considering the consistency with the numerical
result.\cite{Misguich}
\\

(b) \underline{$\sqrt{3}\times\sqrt{3}$ bond ordering patterns}
\\

Here we focus on all possible bond ordering patterns which
are compatible with the $\sqrt{3}\times\sqrt{3}$ enlarged unit cell.
Labeling the 27 links inside the $\sqrt{3}\times\sqrt{3}$ unit cell by $| l \rangle$, a bond
ordering pattern is given by the linear combination of the $| l \rangle$ as,
\begin{align}\label{eqn:bondorder}
| \text{bond order} \rangle \propto & \sum_{l} c_{l} |l\rangle,
\end{align}
where $c_{l}$ is proportional to the amplitude of the singlet correlation
on the link $l$, i.e., $c_{l}$$\propto$$\chi_{l}$.
The vector space spanned by $| l \rangle$ constitutes a
reducible representation (defined as $\Gamma_{\text{bond}}$) of the enlarged point group $G_{P,\textbf{b}}$
whose decomposition into the irreducible representations is given by
\begin{align}
\Gamma_{\text{bond}} &= 2 A_{1} \oplus B_{2} \oplus E_{1} \oplus 2 E_{2} \oplus 2 E_{3} \oplus E_{4} \oplus 3 Q. \nonumber
\end{align}
\begin{figure}[t]
\centering
\includegraphics[width=7 cm]{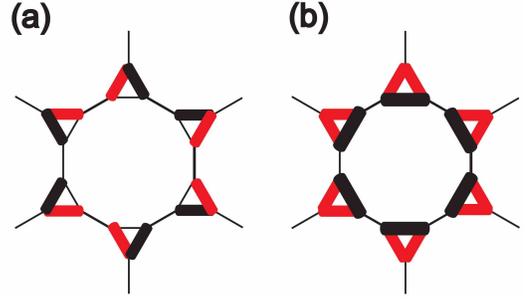}
\caption{(Color online)
Description of the pair of the $E_{3}^{A}$ bond ordered states which form a basis
of the $E_{3}$ irreducible representation.
(a) $c_{l}$=$\sqrt{3}$ for the thick light (red) links, -$\sqrt{3}$
for the thick dark (black) links, and zero for the thin solid links.
(b) $c_{l}$=1 for the thick light (red) links, -2
for the thick dark (black) links, and zero for the thin solid links.
} \label{fig:E3Aorder}
\end{figure}
\begin{figure}[t]
\centering
\includegraphics[width=7 cm]{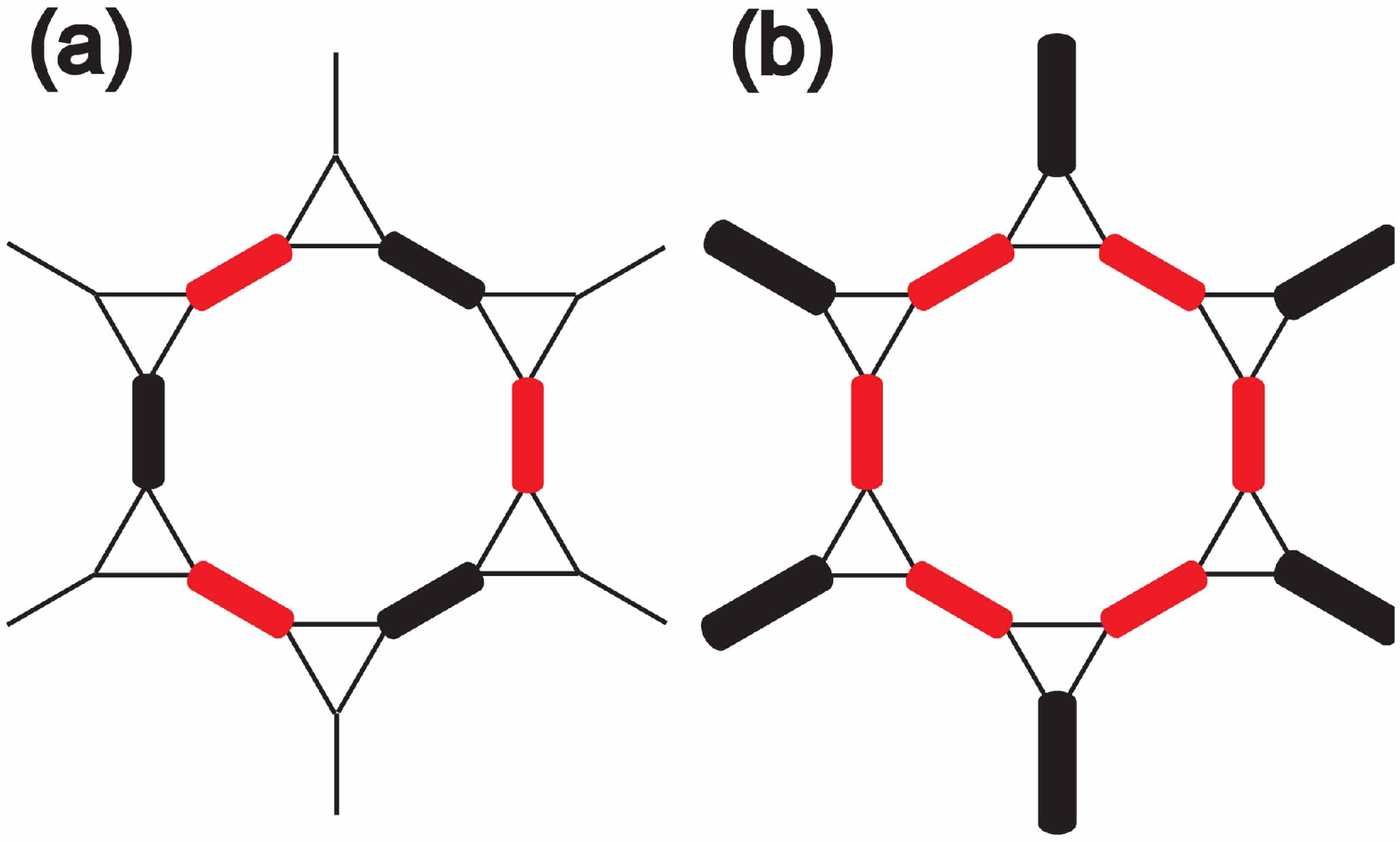}
\caption{(Color online)
Description of the pair of the $E_{3}^{B}$ bond ordered states which form a basis
of the $E_{3}$ irreducible representation.
(a) $c_{l}$=$\sqrt{3}$ for the thick light (red) links, -$\sqrt{3}$
for the thick dark (black) links, and zero for the thin solid links.
(b) $c_{l}$=1 for the thick light (red) links, -2
for the thick dark (black) links, and zero for the thin solid links.
}\label{fig:E3Border}
\end{figure}
\begin{figure}[t]
\centering
\includegraphics[width=7 cm]{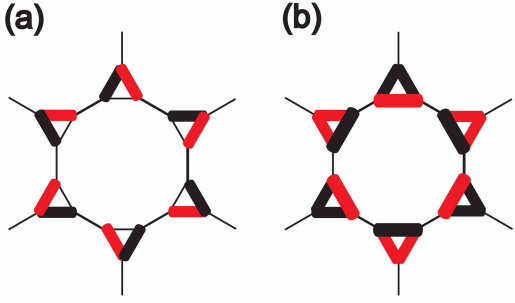}
\caption{(Color online)
Description of the pair of the $E_{4}$ bond ordered states which form a basis
of the $E_{4}$ irreducible representation.
(a) $c_{l}$=$\sqrt{3}$ for the thick light (red) links, -$\sqrt{3}$
for the thick dark (black) links, and zero for the thin solid links.
(b) In each triangle the link belonging to the central dodecagon
has the $c_{l}$=2 (-2) if it has red (black) color. The other two links
of the triangle have $c_{l}$=1 (-1) if they have red (black) colors.
}\label{fig:E4order}
\end{figure}
Notice that $\Gamma_{\text{bond}}$ supports two independent $E_{3}$
irreducible representations (we call them as $E_{3}^{A}$ and $E_{3}^{B}$, respectively)
and one $E_{4}$ irreducible representation.
The bond ordering patterns which constitute a basis of each irreducible
representation are displayed in Fig.~\ref{fig:E3Aorder},
Fig.\ref{fig:E3Border}, and Fig.~\ref{fig:E4order}
describing $E_{3}^{A}$, $E_{3}^{B}$, and $E_{4}$ irreducible
representations, respectively.

A bond order transforming as an $E_{3}$ irreducible representation can be represented by
a linear combination of states like
\begin{align}\label{eqn:bondorderE3}
|E_{3} \text{bond order} \rangle
=&\alpha_{1} |E_{3}^{A}(a)\rangle +
\alpha_{2} |E_{3}^{A}(b)\rangle \nonumber\\
&+\alpha_{3} |E_{3}^{B}(a)\rangle +\alpha_{4} |E_{3}^{B}(b)\rangle,
\end{align}
in which $|E_{3}^{A}(a)\rangle = \sum_{l} c_{l} |l\rangle$ with $c_{l}$
specified in Fig.~\ref{fig:E3Aorder}(a). The other three basis states $|E_{3}^{A}(b)\rangle$,
$|E_{3}^{B}(a)\rangle$, and $|E_{3}^{B}(b)\rangle$ are defined following the same way.

Interestingly the bond ordering patterns of the
two valence bond solid states, the columnar and box 18-site VBS,
are given by the following superposition of states,

\begin{align}\label{eqn:18site-bondorder}
&|\text{Columnar VBS} \rangle
\propto |\text{uniform}\rangle - |E_{3}^{A}(b)\rangle - |E_{3}^{B}(b)\rangle,\nonumber\\
&|\text{Box VBS} \rangle
\propto |\text{uniform}\rangle + \frac{1}{2} |E_{3}^{A}(b)\rangle +\frac{1}{2} |E_{3}^{B}(b)\rangle,
\end{align}
where $|\text{uniform}\rangle \equiv \sum_{l} |l\rangle$.
Since both the columnar and box 18-site VBS are invariant under
the reflections $R_{a}$, $R_{b}$, and $R_{c}$,
$|E_{3}^{A}(a)\rangle$ and $|E_{3}^{B}(a)\rangle$ have no contribution.
Superpositions of the $|E_{3}^{A}(b)\rangle$ and $|E_{3}^{B}(b)\rangle$
can induce more general bond ordering patterns other than those described in Fig.~\ref{fig:Jtdimerconfig}.
Finally, since $E_{4}$ irreducible representation always breaks
the reflections $R_{a}$, $R_{b}$, and $R_{c}$ (See Fig.~\ref{fig:E4order}),
we neglect bond orders transforming as $E_{4}$ irreducible representation.

\section{\label{sec:energetics} Ground state energy}
In this section we compare the ground state energies of various spin liquid states.

\subsection{\label{sec:MFenergy} Mean field theory}
The ground state energies of various spin liquid states
for $J_{t}$=2$J_{e}$ are shown in Table~\ref{table:energy}.
In addition to the translationally invariant spin liquid states with finite $J_{e}$ and $J_{t}$,
we have also considered a decoupled dimer phase for comparison. A decoupled dimer phase,
which has the lowest mean field ground state energy for $J_{t}>J_{e}$, can be built
based on the local dimer configuration described in Fig.~\ref{fig:Jtdimer}(a).
The columnar 18-site valence bond solid (VBS) displayed in Fig.~\ref{fig:Jtdimerconfig}(a) is an example.
According to the mean field calculation,
the dimer state has lower ground state energy than any other translationally invariant spin liquids.
Among the spin liquid phases with translational invariance, the four ansatz having the $\pi/2$ flux inside triangles
have lower energies than those having zero flux inside triangles, i.e., SL$[0,0,0]$ and SL$[0,0,\pi]$.

\begin{table}
\begin{tabular}{@{}|c|c|c|}
\hline \hline
& $E_{MF}$ (unprojected) &  $E_{MF}$ (projected) \\
\hline \hline
\text{Dimer} & -0.625 & -0.625 \\
SL$[0,0,0]$ & -0.498 & -0.647 \\
SL$[\frac{\pi}{2} ,\frac{\pi}{2} , \pi]$ & -0.553 & -0.624 \\
SL$[-\frac{\pi}{2} ,\frac{\pi}{2} , 0]$ & -0.553 & -0.616 \\
SL$[0,0,\pi]$ & -0.498 & -0.654 \\
SL$[\frac{\pi}{2} ,\frac{\pi}{2} , 0]$ & -0.552 & -0.617 \\
SL$[-\frac{\pi}{2} ,\frac{\pi}{2} , \pi]$ & -0.552 & -0.614 \\
\hline \hline
\end{tabular}
\caption{The ground state energies of the various mean field ansatz when $J_{t}$=2$J_{e}$.
The energies are measured in unit of $J_{e}$ }
\label{table:energy}
\end{table}

\begin{figure}[t]
\centering
\includegraphics[width=8 cm]{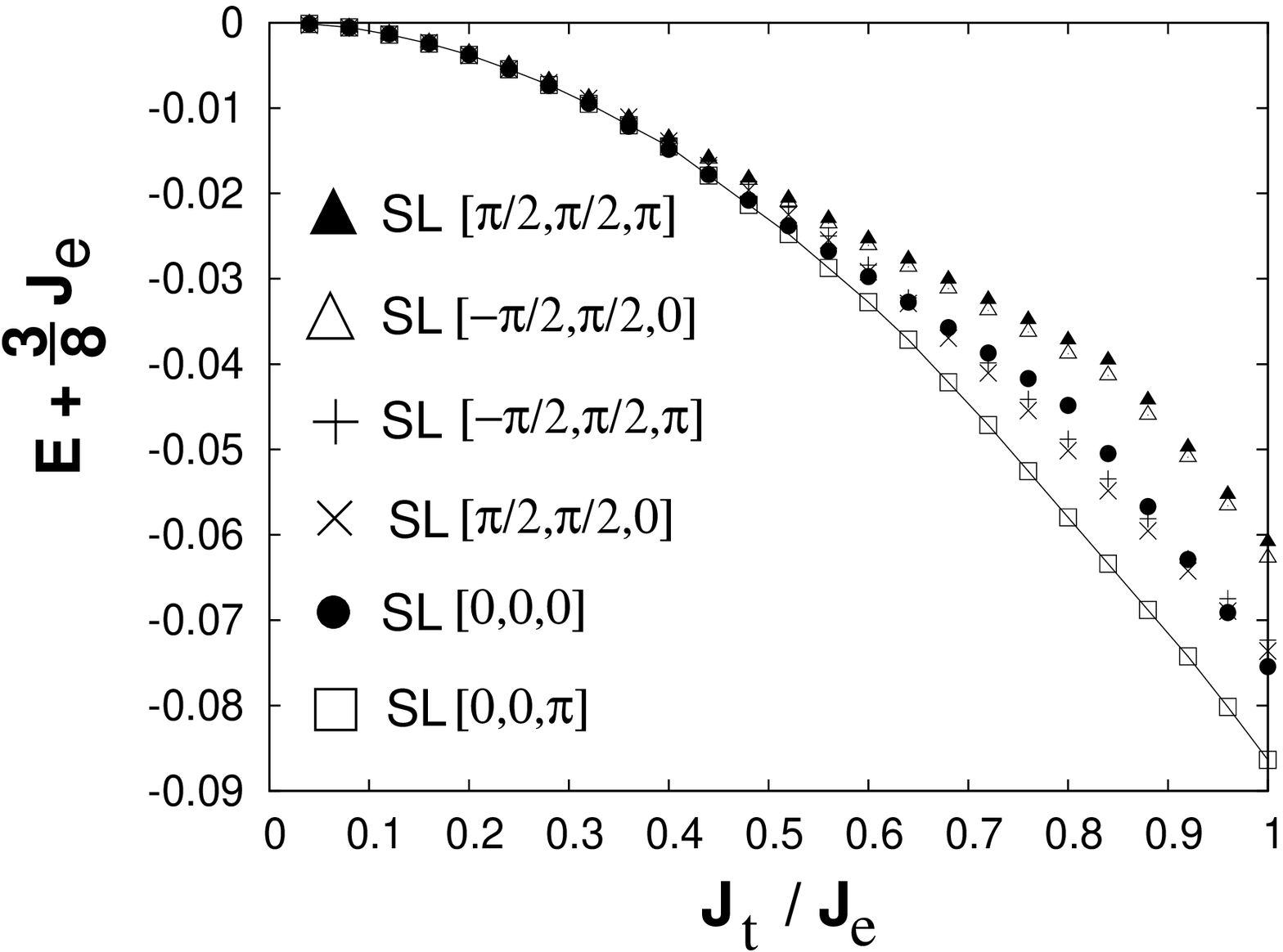}
\includegraphics[width=8 cm]{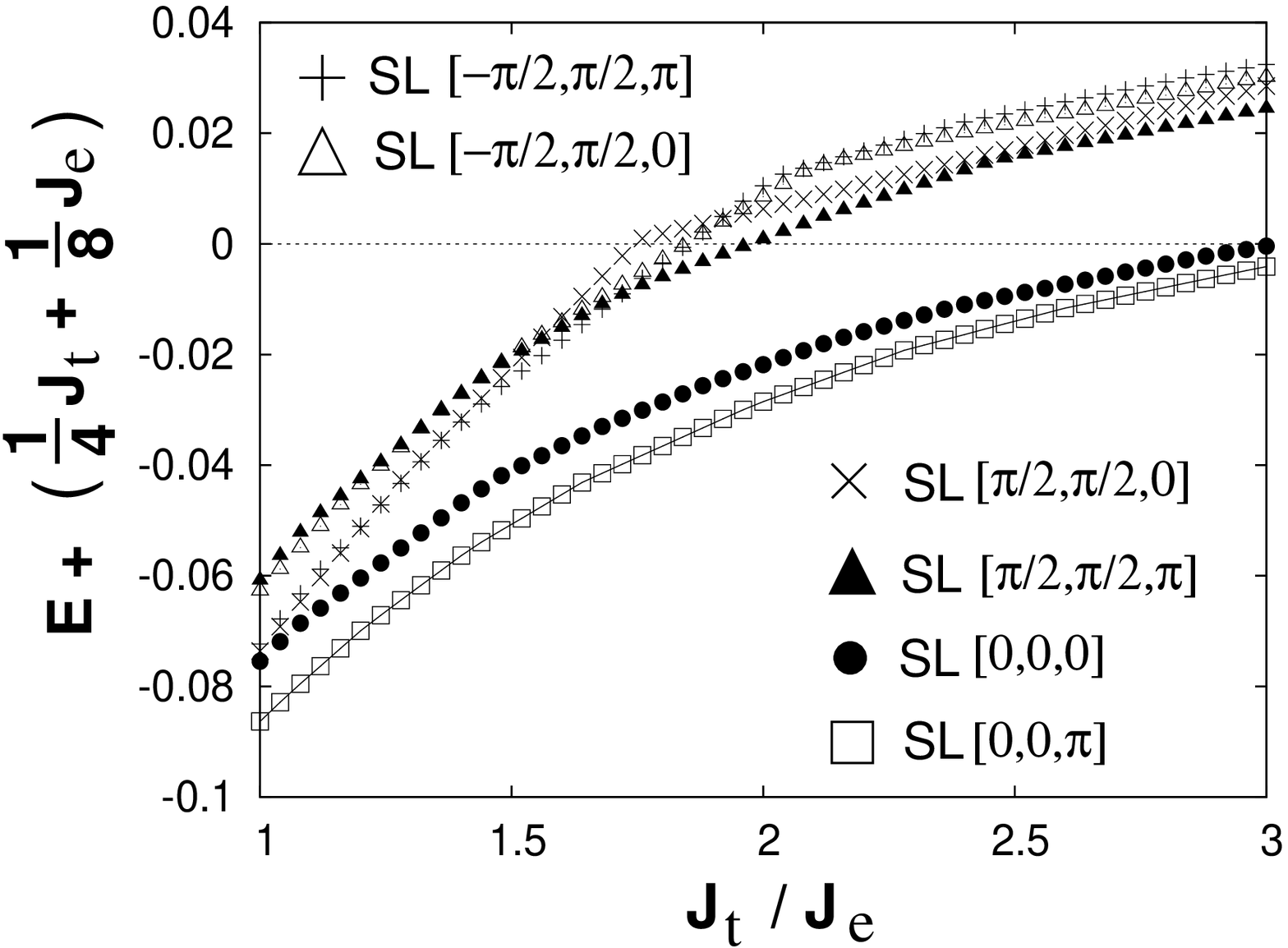}
\caption{(Color online)
The ground state energy per spin of the various spin liquid ansatzes
computed using Gutzwiller projection.
(a) Energies for $0 \leq J_t/J_e \leq 1$ measured
relative to the $J_e$-dimer VBS state shown in Fig.\ref{fig:Jedimer} which has an energy $-3 J_e/8$
per spin.
(b) Energies for $1 \leq J_t/J_e \leq 3$ measured
relative to the $\sqrt{3}\times\sqrt{3}$ VBS state shown in Fig.\ref{fig:Jtdimerconfig} which has an energy
$-J_t/4-J_e/8$ per spin. Note that SL[0,0,$\pi$] (solid line is a guide to the eyes)
has the lowest energy
over the whole parameter space. SL[0,0,0] has the second lowest energy
for a wide range for $J_t > J_e$.  For $J_{t}\!=\!2 J_{e}\!=\!2$,
the energy per spin of all six spin liquid ansatzes
are given in Table.\ref{table:energy}. (These computations were carried out on a system with
$6 \times 6$ unit cells, i.e., with $216$ spins, and the statistical error bars on the energy are of the
order of the symbol size).
} \label{fig:energy_wfn}
\end{figure}

\subsection{\label{sec:VMC} Projected wave function study}
In the above mean field calculation, the single occupancy constraint is
imposed only on average. Therefore the mean field wave functions
contain unphysical states with zero or two fermions at a point. To obtain physical spin wave functions
we therefore perform a numerical Gutzwiller projection on the mean field wave functions.
The ground state energies of the projected states are computed numerically
using the variational Monte Carlo (VMC) method\cite{VMC1,VMC2}.
The resulting energies are displayed in Fig.\ref{fig:energy_wfn} for a range of $J_t/J_e$
where we have optimized the state with respect to $\chi_t/\chi_e$ for each value of $J_t/J_e$.
Table~\ref{table:energy} shows the numerical energy values for $J_t/J_e$=2.0 to facilitate a
comparison with the mean field numerics. We find that Gutzwiller projection dramatically
changes the relative ordering of the various states, and that the state SL$[0,0,\pi]$
appears, upon projection, to be the lowest energy spin liquid over the entire parameter range.

\subsection{\label{sec:bondoperator} Bond operator approach}

According to the projected wave function study, SL[0,0,$\pi$] state
is the ground state over a wide parameter space.
However, the SL[0,0,$\pi$] state is unstable
due to the confinement in the 2+1 dimensional pure gauge theory.
It is also inconsistent with our expectation for the $J_{e}\gg J_{t}$ limit.
When $J_{e}\gg J_{t}$, the $J_e$-dimer VBS phase (See Fig.~\ref{fig:Jedimer}.) is the exact ground state.
In addition the recent exact diagonalization study shows that the $J_e$-dimer VBS phase
remains as the ground state up to the isotropic limit of $J_{e} = J_{t}$.\cite{Richter}
Therefore SL[0,0,$\pi$] state should have higher energy than $J_e$-dimer VBS phase
at least in some finite range of  $0\leq J_{t}/J_{e} \leq 1$.
This discrepancy comes from the lack of the interdimer interaction
in the decoupled dimer limit.
For the description of dimerized phases beyond the decoupled dimer limit,
we undertake the self-consistent bond operator approach.\cite{chubukov,sachdev, Gopalan}
If the correction coming from the inter-dimer interaction
is significant, we also have to check the possibility
that the true ground state is a valence bond solid
even when $J_{t}>J_{e}$.

In the bond operator formulation, the dimer singlet degrees of freedom
are used as natural building blocks and the quantum
corrections coming from the triplet fluctuations
can systematically be investigated.\cite{Kotov1,Kotov2}
Here we present a brief explanation of the bond operator formulation.
Let us consider the two $S=\frac{1}{2}$ spins constituting
a dimer singlet, ${\bf S}_R$
and ${\bf S}_L$. The Hilbert space is spanned by four states
that can be taken as a singlet state, $|s\rangle$, and three
triplet states, $|t_{x}\rangle$, $|t_{y}\rangle$ and
$|t_{z}\rangle$. Then, the singlet and triplet boson operators are
introduced such that each of the above states can be created from
the vacuum $|0\rangle$ as follows:
\begin{align}
|s\rangle  &=s^{\dagger} |0\rangle  =\frac{1}{\sqrt{2}}
(|\uparrow\downarrow\rangle-|\downarrow\uparrow\rangle ),
\nonumber\\
|t_{x}\rangle  &=t_{x}^{\dagger} |0\rangle  =-\frac{1}{\sqrt{2}}
(|\uparrow\uparrow\rangle-|\downarrow\downarrow\rangle ),
\nonumber\\
|t_{y}\rangle  &=t_{y}^{\dagger} |0\rangle  =\frac{i}{\sqrt{2}}
(|\uparrow\uparrow\rangle+|\downarrow\downarrow\rangle ),
\nonumber\\
|t_{z}\rangle  &=t_{z}^{\dagger} |0\rangle  =\frac{1}{\sqrt{2}}
(|\uparrow\downarrow\rangle+|\downarrow\uparrow\rangle ).
\nonumber
\end{align}
To eliminate unphysical states from the enlarged Hilbert space,
the following constraint needs to be imposed on the bond-particle
Hilbert space:
\begin{equation} \label{eq:constraint}
s^{\dagger}s +t_{\alpha}^{\dagger}t_{\alpha} = 1,
\nonumber
\end{equation}
where $\alpha=x,y,$ and $z$, and we adopt the summation convention
for the repeated indices hereafter unless mentioned otherwise.

Constrained by this equation,
the exact expressions for the spin operators can be written in
terms of the bond operators:
\begin{align}\label{eq:bond-ops}
S_{R\alpha}&=\frac{1}{2}(s^{\dag}t_{\alpha} +t_{\alpha}^{\dag}s
-i\varepsilon_{\alpha\beta\gamma}t_{\beta}^{\dag}t_{\gamma}),
\nonumber\\
S_{L\alpha}&=\frac{1}{2}(-s^{\dag}t_{\alpha} -t_{\alpha}^{\dag}s
-i\varepsilon_{\alpha\beta\gamma}t_{\beta}^{\dag}t_{\gamma}),
\nonumber
\end{align}
where $\varepsilon_{\alpha\beta\gamma}$ is the third-rank
totally antisymmetric tensor with $\varepsilon_{xyz}=1$.

Utilizing the bond operator representation of spin
operators, the Heisenberg spin Hamiltonian in Eq.~(\ref{eqn:hamiltonian})
can be rewritten solely in terms of bond
particle operators.
Since all dimers of $J_e$-dimer VBS phase are symmetry
equivalent,
the singlet condensate density
$\langle s_{\textbf{i}} \rangle$ and the chemical potential $\mu_{\textbf{i}}$
can be set to be $\langle s_{\textbf{i}} \rangle$ = $\bar{s}$ and
$\mu_{\textbf{i}}$=$\mu$ in our mean field theory.
Here ${\bf i}$ denotes the
location of dimers.
The hard-core constraint on the bond-particle operators is imposed by
adding the following Lagrange multiplier term,
$H_{\mu} = -\sum_{{\bf i}} \mu (\bar{s}^2 + t^{\dagger}_{{\bf i}\alpha}t_{{\bf i}\alpha}-1)$.
The resuling Hamiltonian can be written as
follows:
\begin{equation}
H=N\epsilon_{0} +H_\textrm{Quad}+H_\textrm{Quartic} , \label{H_total}
\end{equation}
where
\begin{align}
H_\textrm{Quad}=&\left(\frac{J_{e}}{4}-\mu\right)\sum_{\textbf{i}}
t^{\dag}_{\textbf{i}\alpha}t_{\textbf{i}\alpha}
\nonumber\\
&+\frac{{J_{t}\bar{s}^2}}{4}\sum_{\langle\textbf{i},\textbf{j}\rangle} \Big\{
t^{\dag}_{\textbf{i}\alpha}t_{\textbf{j}\alpha}+t_{\textbf{i}\alpha}t_{\textbf{j}\alpha}+\textrm{H.
c.} \Big\},
\end{align}
and
\begin{align}
H_\textrm{Quartic}=
-\frac{{J_{t}}}{4}\sum_{\langle\textbf{i},\textbf{j}\rangle} \varepsilon_{\alpha\beta\gamma}\varepsilon_{\alpha\mu\nu}
t^{\dag}_{\textbf{i}\beta}t_{\textbf{i}\gamma}t^{\dag}_{\textbf{j}\mu}t_{\textbf{j}\nu}.
\end{align}
In the above, $N$ is the number of
unit cells and
\begin{align}
\epsilon_{0}
&=3\left[\mu(1-{\bar{s}}^{2})-\frac{3}{4}J_{e}{\bar{s}}^{2}\right].
\end{align}

\begin{figure}[t]
\centering
\includegraphics[width=7 cm]{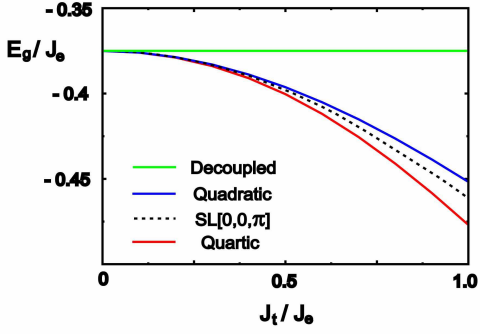}
\caption{(Color online)
Comparison of the ground state energy of SL[0,0,$\pi$]
to that of $J_{e}$-dimer VBS for 0$\leq J_{t}/J_{e} \leq1$.
To emphasize the importance of the interdimer interactions
we present the energies of $J_{e}$-dimer VBS obtained in three different
ways. The 'Decoupled' indicates the energy of the decoupled dimers.
'Quartic' ('Quadratic') is the energy from the bond operator
theory with (without) the quartic interaction effect.
} \label{fig:Energy_Jedimer}
\end{figure}

The quartic interactions between triplet particles are decoupled using
the mean field order parameters $P$ and $Q$, where
$P \equiv \langle t^{\dagger}_{{\bf i}\alpha}t_{{\bf j}\alpha}  \rangle$
and $Q \equiv \langle t_{{\bf i}\alpha}t_{{\bf j}\alpha}  \rangle$.
Here $P$ and $Q$ denote
the diagonal and off-diagonal triplet correlations between neighboring dimers.
These two order parameters $P$ and $Q$
together with $\bar{s}$ and $\mu$ are determined self-consistently
by solving the coupled saddle point equations.\cite{sachdev,Gopalan}

The ground state energy of $J_e$-dimer VBS phase obtained
from the self-consistent bond operator calculation is displayed
and compared to the energy of SL[0,0,$\pi$] state in Fig.~\ref{fig:Energy_Jedimer}.
Here we have obtained the energy of $J_e$-dimer VBS phase in three different ways.
If we neglect the inter-dimer couplings (the decoupled dimer limit) completely,
the energy is independent of $J_t$/$J_e$.
The inclusion of the inter-dimer interaction
lowers the ground state energy significantly. In the end,
$J_e$-dimer VBS phase has the lower ground energy
than the SL[0,0,$\pi$] state over the entire
parameter range of 0$\leq$$J_t$/$J_e$$\leq$1
when we include the quartic interactions.
The inter-dimer interactions generate
huge correction to the ground state energy
of dimerized phases.

\begin{figure}[t]
\centering
\includegraphics[width=8.5 cm]{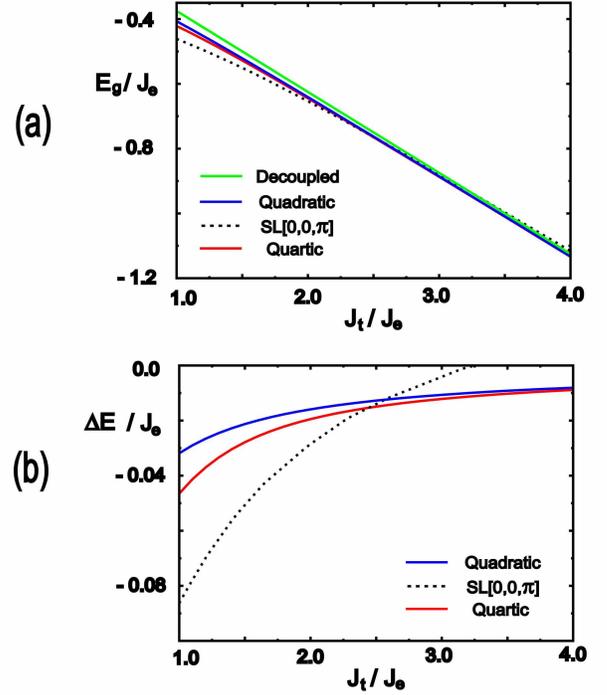}
\caption{(Color online)
Comparison of the ground state energies of the columnar 18-site VBS
to that of SL[0,0,$\pi$] state.
(a) Energetics for $J_{t}$$>$$J_{e}$. The energies of the columnar 18-site VBS
are obtained in three different ways as in Fig.\ref{fig:Energy_Jedimer}.
(b) The relative energies obtained by subtracting the decoupled dimer energy.
Note that
there is a level crossing around $J_{t}$/$J_{e}$$\approx$2.4
between the energy of SL[0,0,$\pi$] state and that of the columnar 18-site VBS
including the quartic interaction.
} \label{fig:00pi_dimer}
\end{figure}

Now we concentrate on the other limit where $J_t$$>$$J_e$.
In contrast to the $J_e$$\gg$$J_t$ limit, it is nontrivial
to identify the ground state even when we restrict our
attention to valence bond solid phases. Taking into account
the information from the exact diagonalization study
and 1/N fluctuation from the large-N limit, we suggest
the columnar 18-site VBS phase as a promising candidate for the ground state
as explained below.

We apply the bond operator approach to the columnar 18-site VBS phase.
The nine dimers within the unit cell can be divided into two groups.
One group is made of the six dimers lying on the triangular links.
Note that all these six dimers are lying on a dodecagon.
(See the central dodecagon in Fig.~\ref{fig:Jtdimerconfig}(a).)
We call such a dodecagon surrounded by six dimers as a ``perfect" dodecagon.
The remaining three dimers lying on the expanded links
make the other group. Every dimer belonging
to the same group is symmetry equivalent as one can
easily notice from the patterns around the central ``perfect" dodecagon
in Fig.~\ref{fig:Jtdimerconfig}(a). To apply the bond operator
approach we have to introduce two independent sets of order parameters
to distinguish the two different groups of dimers.
We use $\bar{s}_{e}$ and $\mu_{e}$ ($\bar{s}_{t}$ and $\mu_{t}$)
to indicate the singlet condensate density and the chemical potential
corresponding to the expanded (triangular) link.
To decouple the quartic triplet interactions we introduce
two sets of the order parameters, that is, $\{$$P_{pp}$, $Q_{pp}$$\}$
and $\{$$P_{ep}$, $Q_{ep}$$\}$. $P_{pp}$ ($Q_{pp}$) describes the diagonal
(off-diagonal) correlation between the neighboring dimers
lying on a ``perfect" dodecagon. On the other hand
$P_{ep}$ ($Q_{ep}$) describes the diagonal
(off-diagonal) correlation between a dimer lying on an expanded link
and its neighboring dimer lying on a ``perfect" dodecagon.
We have determined the eight parameters $\bar{s}_\alpha$, $\mu_\alpha$ ($\alpha$ = t or e),
and $P_\beta$, $Q_\beta$ ($\beta$ = $pp$ or $ep$) self-consistently
by solving the coupled saddle point equations.

The self-consistent solution shows that $P_{ep}$=$Q_{ep}$=0, ${\bar{s}_e}^2$=1 and $\mu_e$=-3/4$J_{e}$.
Since $P_{ep}$ and $Q_{ep}$ describe the coupling between the dimers lying on expanded links
and the dimers lying on ``perfect" dodecagons, these two groups of dimers
are completely decoupled when $P_{ep}$=$Q_{ep}$=0.
In this situation, every dimer lying on expanded links
is decoupled from the surrounding, leading to ${\bar{s}_e}^2$=1 and $\mu_e$=-3/4$J_{e}$.
The triplet fluctuations are confined inside every isolated perfect dodecagon,
which is reflected in the finite $P_{pp}$ and $Q_{pp}$ values. This interesting structure
would result in the highly localized triplet excitation spectrum.

The ground state energies of the columnar 18-site VBS and SL[0,0,$\pi$]
are compared in Fig.~\ref{fig:00pi_dimer}.
The energies of the columnar 18-site VBS are obtained
in three different ways again.
That is, for the decoupled dimer limit, including the inter-dimer coupling neglecting
quartic interactions, and finally including the inter-dimer quartic interactions.
For clarity we also calculated the energy difference
relative to the decoupled dimer energy as shown in Fig.~\ref{fig:00pi_dimer}(b).
Interestingly,
there is a critical ratio $(J_t/J_e)_c$$\approx$ 2.4 beyond which
the columnar 18-site VBS becomes the ground state when we include the quartic triplet interactions.
Even though the critical ratio $(J_t/J_e)_c$ is a bit larger
than the suggested phase boundary from the numerical study\cite{Misguich},
the existence of the critical values of $(J_t/J_e)_c$ is quite encouraging.
In particular, because the slopes of the lines in Fig.~\ref{fig:00pi_dimer}(a)
are almost parallel, small additional energy correction
could induce a large shift of the crossing point as shown in Fig.~\ref{fig:00pi_dimer}(b).
Since the simple Hartree-Fock approximation does not
take into account
the fluctuations coming from the cooperative interaction
between the dimers on the expanded links and those
on the perfect dodecagons, we expect that the quantum correction
beyond the Hartree-Fock limit could shift the energy
level crossing point down to $(J_t/J_e)_c$$\approx$1.3
as suggested by the numerical study.\cite{Misguich}

\section{\label{sec:sl00pi} Instability of SL[0,0,$\pi$] spin liquid and valence bond solids}
Summarizing the previous discussions, $J_e$-dimer VBS phase is the ground state for $J_{t}/J_{e}<(J_{t}/J_{e})_{c1}$
while the columnar 18-site VBS is the ground state in the opposite limit of $J_{t}/J_{e}>(J_{t}/J_{e})_{c2}$.
(Here $(J_{t}/J_{e})_{c1}\leq(J_{t}/J_{e})_{c2}$.)
Although this result is obtained based on the energy comparison with SL[0,0,$\pi$] state,
one may still expect that the two valence bond solid states are intimately related to the SL[0,0,$\pi$] state.
In particular, the $J_e$-dimer VBS
and the columnar 18-site VBS may arise as a consequence of the confinement in the SL[0,0,$\pi$] spin liquid state.
In this section we describe the possible relation between these two valence bond solid phases
and the SL[0,0,$\pi$] state.

\subsection{\label{sec:monopole} Spinon confinement and uniform bond orders}

Due to the finite spinon gap, the U(1)
gauge field is the only low energy excitation in the SL[0,0,$\pi$] state in the long wavelength limit.
Since the compact U(1) gauge theory without matter field is confining in 2+1 dimension,
we expect that the monopole proliferation would lead the SL$[0,0,\pi]$ ansatz
to some confined phases. To understand the properties of the confined phases
resulting from the monopole condensation,
we have to determine the symmetry properties of the monopole operators.

Here we discuss the possibility that
the monopole operators are invariant under all possible symmetry transformations.
The $J_e$-dimer VBS phase, which is the ground state for $J_{t}/J_{e}<1$,
is invariant under space group operations.
The bond ordering pattern of the $J_e$-dimer VBS phase
belongs to the trivial $A_{1}$ irreducible representation
of the $D_{6}$ point group.
Therefore if we interpret the $J_{e}$-dimer VBS phase
to be induced by the confinement transition,
which is reasonable in the limit of $J_{t}/J_{e}\ll 1$,
this reflects the fact that monopole operators
are invariant under symmetry transformations.

\begin{figure}[t]
\centering
\includegraphics[width=7 cm]{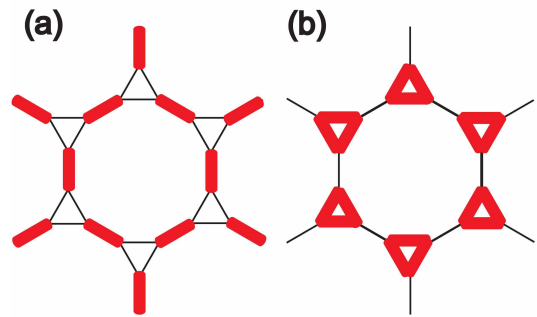}
\caption{(Color online)
Two bond ordering patterns transforming as the $A_{1}$ irreducible
representation. The thick solid (red) link has $c_{l}$ = 1 while
the thin solid link has $c_{l}$=0.
(a) $J_e$-bond order.
(b) $J_t$-bond order.
} \label{fig:A1order}
\end{figure}

Extending the group theory
analysis we performed in Sec.~\ref{sec:grouptheory},
we investigate all possible bond orders invariant under the space group operations.
These are displayed in Fig.~\ref{fig:A1order}.
Here we have finite singlet correlation ($c_{l} \neq 0$) only on the thick solid (red) links.
The bond order in Fig.~\ref{fig:A1order}(a) is nothing but the
$J_{e}$-dimer VBS phase. On the other hand, the bond order
in Fig.~\ref{fig:A1order}(b) has finite $c_{l}$ only on the triangular
links (we call it as a $J_{t}$-bond ordered phase). We expect the $J_{t}$-bond
ordered phase is the natural low energy bond ordering pattern when $J_{t}$$\gg$$J_{e}$.
Since the arbitrary superposition of these two orders follows
the same $A_{1}$ irreducible representation, we expect the actual ground states
would have finite $c_{l}$ values over all the links on the lattice.
However, it is natural to expect that the $c_{l}$ on the expanded (triangular) link
would be larger than that on the triangular (expanded) link when $J_{e}$ $>$ $J_{t}$
($J_{e}$ $<$ $J_{t}$). Therefore the bond ordered phase corresponding
to the $A_{1}$ irreducible representation successfully describes the low energy manifold
over the whole parameter range of $J_{t}$/$J_{e}$.
Interestingly, the recent work by Choy and Kim\cite{choy}
has suggested that the same bond ordered states are the ground states
of the same model Hamiltonian
in the strong quantum limit based on the bosonic Sp(N) approach.

Since the spinon bandstructure of the SL[0,0,$\pi$] state
does not change qualitatively by varying $J_{t}$/$J_{e}$,
we expect that the change of $J_{t}$/$J_{e}$ ratio
would not affect the trivial monopole quantum number. Therefore
if the ordered phase is coming from the confinement transition,
it will transform trivially under the symmetry operations.
However, it is also possible that the instability
of the SL[0,0,$\pi$] state is caused by the
interactions between spinons leading to some broken
symmetry phases. We discuss about this possibility
in the following section.


\subsection{\label{sec:psg} Instability induced by interactions between spinons}

Here we investigate the instability of the SL[0,0,$\pi$] state
coming from the interactions between spinons and
the symmetry properties of the resulting ordered phase.
Especially we focus on the instability towards the states
with the  $\sqrt{3}\times\sqrt{3}$-type translational symmetry breaking.
As shown in Fig.~\ref{fig:band00pi}, the valence band is completely flat
without any preferred momentum. However, the conduction band supports
several dispersion minima.
The four minimum points of the conduction band are given by
\begin{align}
\textbf{m}_{3} =& -\textbf{m}_{2} = (\frac{\pi}{6}, 0)\quad\text{and}\quad
\textbf{m}_{4} = -\textbf{m}_{1} = (\frac{\pi}{3},\frac{\pi}{2\sqrt{3}}).\nonumber
\end{align}
Interestingly if we double the vectors connecting neighboring minimum points,
they sit on the Brillouin zone corners which are nothing but the momentum
corresponding to the $\sqrt{3}\times\sqrt{3}$ ordering. Motivated by this
observation we study the symmetry properties of the bound states made of
low energy fermions near the conduction band minima.

We introduce the fermion fields
$\Psi_{i}$ which describe the low energy excitations near the four conduction
band minima $\textbf{m}_{i}$ ($i$ = 1,2,3 and 4),

\begin{align}
\Psi_{i}(\textbf{x}) &\sim \sum_{n=1}^{12}e^{-i\textbf{m}_{i}\cdot\textbf{x}}(\nu_{i})^{*}_{n} f_{n}(\textbf{x}),\nonumber
\end{align}
where $\nu_{i}$ is the eigenvector of the mean field Hamiltonian at the momentum $\textbf{m}_{i}$,
and $f_{n}$ is a slowly varying fermion field near the conduction band minimum, with $n$
labelling the twelve sites within the unit cell.
To determine the symmetry of bound states made of the above low energy fermions,
we have to understand how the symmetries of the microscopic Hamiltonian
are realized in the effective continuum fields, $\Psi_{i}$.
Here we follow the same procedure which we use
to determine the transformation properties of the continuum
field for the SL$[-\frac{\pi}{2} ,\frac{\pi}{2} , 0]$ state in Sec.~\ref{sec:properties}.
To extract the necessary information on the transformations
of the continuum field, we consider a finite system
of a 6$\times$12 unit cell. By solving the mean field
Hamiltonian on this finite system, we determine the properties of the eigenvectors
at the four momentum points, $\textbf{m}_{i}$.
The detailed explanation of the procedure as to how to determine the symmetry
of the continuum fields is discussed in the Appendix~\ref{sec:fieldsymmetry}.

Through the projective symmetry group analysis, we determine the following
transformation properties of the continuum fields,
\begin{displaymath}
T_{a_{1}} :
\left( \begin{array}{cccc}
\Psi_{1}  \\
\Psi_{3}  \\
\Psi_{2}  \\
\Psi_{4}
\end{array} \right)
\rightarrow
\left( \begin{array}{cccc}
0 & e^{-i\frac{\pi}{3}} & 0 & 0 \\
1 & 0 & 0 & 0 \\
0 & 0 & 0 & 1 \\
0 & 0 & e^{i\frac{\pi}{3}} & 0
\end{array} \right)
\left( \begin{array}{cccc}
\Psi_{1}  \\
\Psi_{3}  \\
\Psi_{2}  \\
\Psi_{4}
\end{array} \right),
\end{displaymath}

\begin{displaymath}
T_{a_{2}} :
\left( \begin{array}{cccc}
\Psi_{1}  \\
\Psi_{3}  \\
\Psi_{2}  \\
\Psi_{4}
\end{array} \right)
\rightarrow
\left( \begin{array}{cccc}
e^{-i\frac{5\pi}{6}} & 0 & 0 & 0 \\
0 & e^{i\frac{\pi}{6}} & 0 & 0 \\
0 & 0 & e^{i\frac{\pi}{6}} & 0\\
0 & 0 & 0 & e^{i\frac{5\pi}{6}}
\end{array} \right)
\left( \begin{array}{cccc}
\Psi_{1}  \\
\Psi_{3}  \\
\Psi_{2}  \\
\Psi_{4}
\end{array} \right),
\end{displaymath}

\begin{displaymath}
R_{y} :
\left( \begin{array}{cccc}
\Psi_{1}  \\
\Psi_{3}  \\
\Psi_{2}  \\
\Psi_{4}
\end{array} \right)
\rightarrow
\left( \begin{array}{cccc}
\frac{-1}{\sqrt{2}}&\frac{-e^{i\frac{\pi}{3}}}{\sqrt{2}}&0&0\\
\frac{-e^{-i\frac{\pi}{3}}}{\sqrt{2}}&\frac{1}{\sqrt{2}}&0&0\\
0&0&\frac{1}{\sqrt{2}}&\frac{-e^{i\frac{\pi}{3}}}{\sqrt{2}}\\
0&0&\frac{-e^{-i\frac{\pi}{3}}}{\sqrt{2}}&\frac{-1}{\sqrt{2}}
\end{array} \right)
\left( \begin{array}{cccc}
\Psi_{1}  \\
\Psi_{3}  \\
\Psi_{2}  \\
\Psi_{4}
\end{array} \right),
\end{displaymath}

\begin{displaymath}
C_{\frac{\pi}{3}} :
\left( \begin{array}{cccc}
\Psi_{1}  \\
\Psi_{3}  \\
\Psi_{2}  \\
\Psi_{4}
\end{array} \right)
\rightarrow
\left( \begin{array}{cccc}
0 & 0 & \frac{-1}{\sqrt{2}} & \frac{-e^{i\frac{\pi}{3}}}{\sqrt{2}}\\
0 & 0 & \frac{-ie^{-i\frac{\pi}{3}}}{\sqrt{2}}&\frac{i}{\sqrt{2}}\\
\frac{-i}{\sqrt{2}} & \frac{ie^{i\frac{\pi}{3}}}{\sqrt{2}} & 0 & 0\\
\frac{-e^{-i\frac{\pi}{3}}}{\sqrt{2}} & \frac{-1}{\sqrt{2}} & 0 & 0
\end{array} \right)
\left( \begin{array}{cccc}
\Psi_{1}  \\
\Psi_{3}  \\
\Psi_{2}  \\
\Psi_{4}
\end{array} \right).
\end{displaymath}


We first investigate the symmetry of all possible fermion bilinears
which can be written as $\Psi_{i}^{\dag}\text{M}_{ij}\Psi_{j}$.
For the description of the 4 by 4 unitary matrix $\text{M}_{ij}$
we introduce two sets of the pauli matrices $\tau_{i}$ and $\mu_{j}$.
Here $\tau_{i}$ is acting on the space spanned by either ($\Psi_{1}$,$\Psi_{3}$)
or ($\Psi_{2}$,$\Psi_{4}$). On the other hand $\mu_{i}$ is defined in
the space spanned by ($\Psi_{1}$,$\Psi_{2}$) or ($\Psi_{3}$,$\Psi_{4}$).
Among the sixteen possible bilinears, there are only two terms which
can form a basis of the enlarged point group $G_{P,\textbf{b}}$.
These are $\Psi_{i}^{\dag}\tau_{3}\mu_{0}\Psi_{j}$ and $\Psi_{i}^{\dag}\tau_{0}\mu_{0}\Psi_{j}$
which transform as the $A_{2}$ and $A_{1}$ irreducible representation of the
enlarged point group $G_{P,\textbf{b}}$, respectively. However, since these two bilinears
have zero total momentum, they cannot describe the $\sqrt{3}\times\sqrt{3}$ type
symmetry breaking.
The net momentums carried by the other fourteen bilinears are
neither zero nor $\textbf{K}$=($\frac{2\pi}{3}$,0) . Therefore the transformation
properties of them are not compatible with the symmetry of the
$\sqrt{3}\times\sqrt{3}$ type enlarged unit cell.

Next we consider the instability in the particle-particle channels.
Defining the pairing amplitude as $\Delta_{ij}$$\equiv$$\Psi_{i}\Psi_{j}$,
we have sixteen different $\Delta_{ij}$. We first omit the indices for the spin degrees
of freedom and study how they transform under the space group symmetry operations.
It can easily be checked that the sixteen pairing amplitudes are divided
into the four different sets $\Pi_{1}$, $\Pi_{2}$, $\Omega_{1}$ and $\Omega_{2}$
which transform independently.
The six-component vectors $\Pi_{i}$ and the two-component vectors $\Omega_{i}$
are given by

\begin{displaymath}
\Pi_{1} =
\left( \begin{array}{cccccc}
e^{-i\frac{\pi}{3}}\Delta_{11} + e^{i\frac{\pi}{3}}\Delta_{33}  \\
e^{-i\frac{\pi}{3}}\Delta_{11} - e^{i\frac{\pi}{3}}\Delta_{33}  \\
\Delta_{13}+\Delta_{31}  \\
e^{-i\frac{\pi}{3}}\Delta_{22} + e^{i\frac{\pi}{3}}\Delta_{44}  \\
e^{-i\frac{\pi}{3}}\Delta_{22} - e^{i\frac{\pi}{3}}\Delta_{44}  \\
\Delta_{24}+\Delta_{42}
\end{array} \right),
\end{displaymath}
\begin{displaymath}
\Pi_{2} =
\left( \begin{array}{cccccc}
e^{-i\frac{\pi}{3}}\Delta_{12} + e^{i\frac{\pi}{3}}\Delta_{34}  \\
e^{-i\frac{\pi}{3}}\Delta_{12} - e^{i\frac{\pi}{3}}\Delta_{34}  \\
\Delta_{14}-\Delta_{32}  \\
e^{-i\frac{\pi}{3}}\Delta_{21} + e^{i\frac{\pi}{3}}\Delta_{43}  \\
e^{-i\frac{\pi}{3}}\Delta_{21} - e^{i\frac{\pi}{3}}\Delta_{43}  \\
\Delta_{41}-\Delta_{23}
\end{array} \right),
\end{displaymath}
\begin{displaymath}
\Omega_{1} =
\left( \begin{array}{cc}
\Delta_{13}-\Delta_{31}  \\
\Delta_{24}-\Delta_{42}
\end{array} \right),
\Omega_{2} =
\left( \begin{array}{cccccc}
\Delta_{14}+\Delta_{32}  \\
\Delta_{41}+\Delta_{23}
\end{array} \right).
\end{displaymath}
Under a
space group symmetry operation S, they transform in the
following way,
\begin{align}
\Pi_{i} \rightarrow A_{S}(\Pi_{i}) \Pi_{i}\quad \text{and}\quad\Omega_{i} \rightarrow B_{S}(\Omega_{i})\Omega_{i},\nonumber
\end{align}
where $A_{S}(\Pi_{i})$ ($B_{S}(\Omega_{i})$) is the 6 by 6 (2 by 2) matrix  representing the symmetry operation S,
whose detailed expressions are displayed in the Appendix~\ref{sec:matrices}.

Since the pairing amplitude $\Delta_{ij}$ is not a gauge invariant
object, we have to look into the symmetry of the bilinears such as
$D_{ij,kl}$$\equiv$$\Delta^{\dag}_{ij}$$\Delta_{kl}$.
Even though the number of all possible tensors $D_{ij,kl}$ is very large,
we can reduce the complexity of the symmetry analysis
by focusing on the objects carrying the momentum compatible with
the $\sqrt{3}\times\sqrt{3}$ ordering. This idea leads us to exclude $D_{ij,kl}$ made of
the basis $\Pi_{2}$ and $\Omega_{2}$. In addition, $\Pi_{1}$ and
$\Omega_{1}$ have opposite spin parities, that is, $\Pi_{1}$ is spin singlet
while $\Omega_{1}$ is spin triplet. Therefore all we have to consider are
the terms like $\Pi^{\dag}_{1}$$\text{M}_{\Pi}$$\Pi_{1}$ and
$\Omega^{\dag}_{1}$$\text{M}_{\Omega}$$\Omega_{1}$.

First, we define a set of pauli matrices $\tau_{i}$ acting on the space
spanned by $\Omega_{1}$. Using the transformation properties
of $\Omega_{1}$ under the space group, we obtain a pair
($\Omega^{\dag}_{1}$$\tau_{x}$$\Omega_{1}$,$\Omega^{\dag}_{1}$$\tau_{y}$$\Omega_{1}$)
which has the momentum $\textbf{K}$=($\frac{2\pi}{3}$,0) and
transforms as the $E_{3}$ irreducible representation
of the enlarged point group, $G_{P,\textbf{b}}$.

To understand the symmetry of $\Pi^{\dag}_{1}$$\text{M}_{\Pi}$$\Pi_{1}$,
we introduce a set of the Gell-Mann matrices $\lambda_{a}$ ($a$=1,....,8)\cite{Georgi_book}
as well as the Pauli matrices $\tau_{i}$. For convenience,
we also define the matrix, $\lambda_{9}$=$\sqrt{\frac{2}{3}}$$\mathbf{I}_{3}$
where $\mathbf{I}_{3}$ is the 3$\times$3 identity matrix.
The Gell-Mann matrices are acting on the space spanned by either
($\Pi_{1,1}$, $\Pi_{1,2}$, $\Pi_{1,3}$) or
($\Pi_{1,4}$, $\Pi_{1,5}$, $\Pi_{1,6}$) and $\tau_{i}$
connects these two three-component vectors. Here $\Pi_{1,n}$
indicates the nth component of $\Pi_{1}$.

We have examined the symmetry of all possible bilinears $\Pi^{\dag}_{1}$$\text{M}_{\Pi}$$\Pi_{1}$
and found out that there are only two sets of bilinears which have
the momentum $\textbf{K}$ compatible with the $\sqrt{3}\times\sqrt{3}$ ordering. These are
given by

\begin{displaymath}
\textbf{X}_{E_{3}} =
\left( \begin{array}{cc}
\frac{2\sqrt{2}}{3}\Pi^{\dag}_{1}\tau_{1}\lambda_{8}\Pi_{1}+\frac{1}{3}\Pi^{\dag}_{1}\tau_{1}\lambda_{9}\Pi_{1}  \\
\frac{2\sqrt{2}}{3}\Pi^{\dag}_{1}\tau_{2}\lambda_{8}\Pi_{1}+\frac{1}{3}\Pi^{\dag}_{1}\tau_{2}\lambda_{9}\Pi_{1}
\end{array} \right),
\end{displaymath}
\begin{displaymath}
\textbf{X}_{Q} =
\left( \begin{array}{cccc}
\Pi^{\dag}_{1}\tau_{1}\lambda_{3}\Pi_{1}  \\
\frac{1}{3}\Pi^{\dag}_{1}\tau_{1}\lambda_{8}\Pi_{1}-\frac{2\sqrt{2}}{3}\Pi^{\dag}_{1}\tau_{1}\lambda_{9}\Pi_{1}  \\
\Pi^{\dag}_{1}\tau_{2}\lambda_{3}\Pi_{1}  \\
\frac{1}{3}\Pi^{\dag}_{1}\tau_{2}\lambda_{8}\Pi_{1}-\frac{2\sqrt{2}}{3}\Pi^{\dag}_{1}\tau_{2}\lambda_{9}\Pi_{1}
\end{array} \right).
\end{displaymath}

In the above $\textbf{X}_{E_{3}}$ transforms as the two dimensional $E_{3}$
irreducible representation of the enlarged point group $G_{P,\textbf{b}}$.
On the other hand $\textbf{X}_{Q}$ constitutes
a basis of the four dimensional $Q$ irreducible representation.
Therefore the instability given by $\textbf{X}_{E_{3}}$
has the symmetry consistent with the $\sqrt{3}\times\sqrt{3}$
orders which we have discussed in detail in previous sections.

Since we are considering an instability from
a gapped phase, to stabilize the resulting ordered state,
the condensation energy should be larger than
the excitation gap. However, because the magnitude of the energy gap
reduces as $J_t$/$J_e$ increases, the instability can occur
beyond the critical value of $J_t$/$J_e$.
Finally, since the condition of $\langle\textbf{X}_{E_{3}}\rangle\neq0$ does not
constrain the magnitude of $\langle\Pi_{1}\rangle$,
$\langle\Pi_{1}\rangle$ can have both zero and nonzero values.
If $\langle\Pi_{1}\rangle\neq0$ with $\langle\textbf{X}_{E_{3}}\rangle\neq0$, we have $Z_{2}$ spin liquid
supporting fractionalized quasi-particles and breaking
the translational symmetry at the same time.
This state is similar to the Amperean paired state\cite{ampearean}, which
is recently suggested as a possible ground state of the organic compound $\kappa$-(BEDT-TTF)$_{2}$Cu$_{2}$(CN)$_{3}$.
On the other hand, if $\langle\Pi_{1}\rangle=0$ while $\langle\textbf{X}_{E_{3}}\rangle\neq0$,
we have more conventional phase transforming as an $E_{3}$ irreducible representation.

In conclusion, the symmetry analysis of the low energy fermions
near the conduction band minima shows that the instability of the SL[0,0,$\pi$] state from the particle-particle
channel supports valence bond solid phases,
which have the $\sqrt{3}\times \sqrt{3}$ unit cell
transforming as $E_{3}$ irreducible representations of the enlarged point group $G_{P,\textbf{b}}$.
Remember that the $J_{e}$-dimer VBS phase is induced via the monopole condensation from the SL[0,0,$\pi$] state for small $J_{t}/J_{e}$ limit.
Therefore we can obtain the valence bond solid ground states both for $J_{t}/J_{e}\ll1$ and $J_{t}/J_{e}\gg1$ limits
from the instability of the SL[0,0,$\pi$] state.

\subsection{\label{sec:vmc2} Projected wave function approach}
\begin{figure}[t]
\centering
\includegraphics[width=4.0 cm]{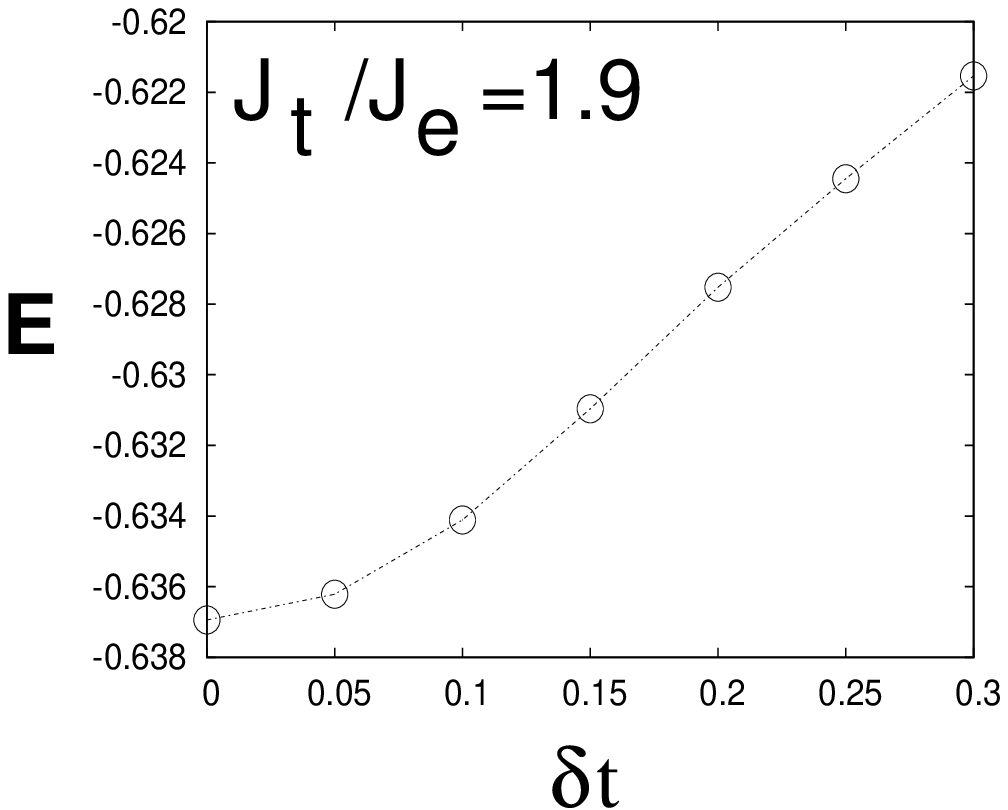}
\includegraphics[width=4.0 cm]{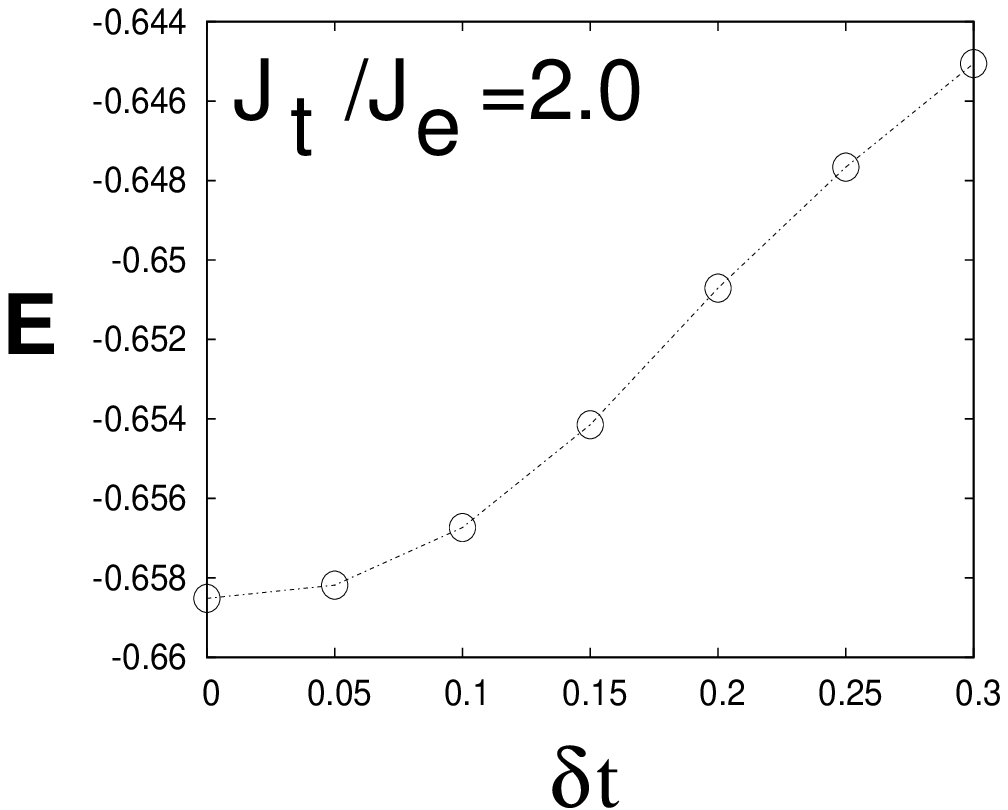}
\includegraphics[width=4.0 cm]{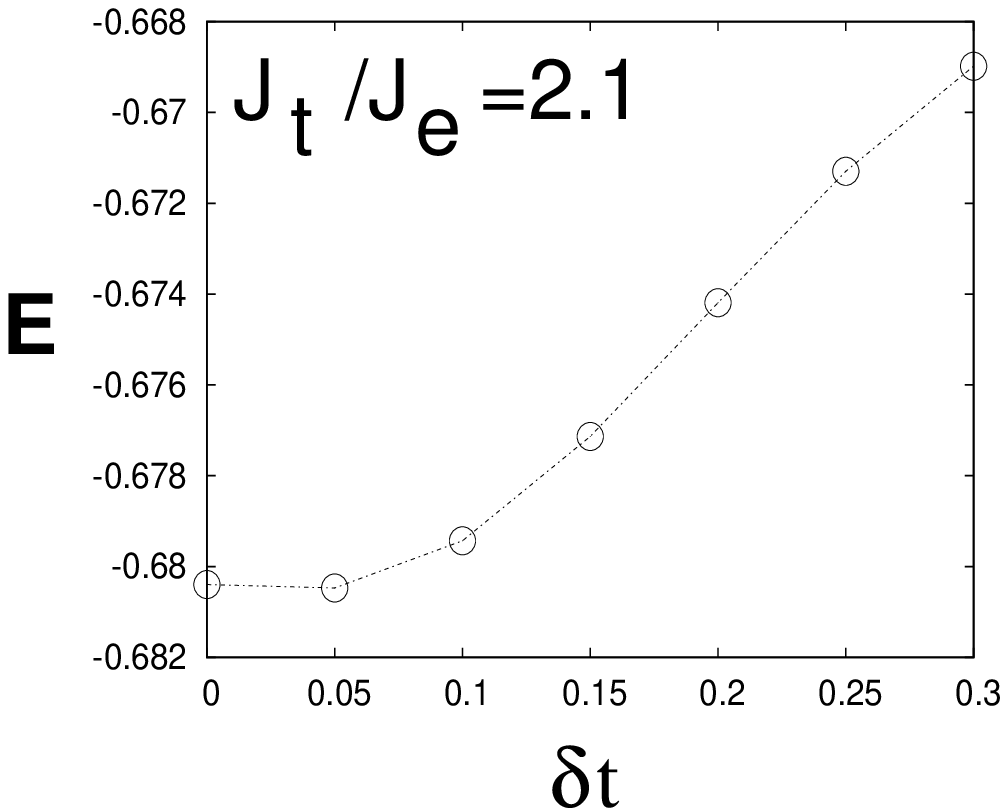}
\includegraphics[width=4.0 cm]{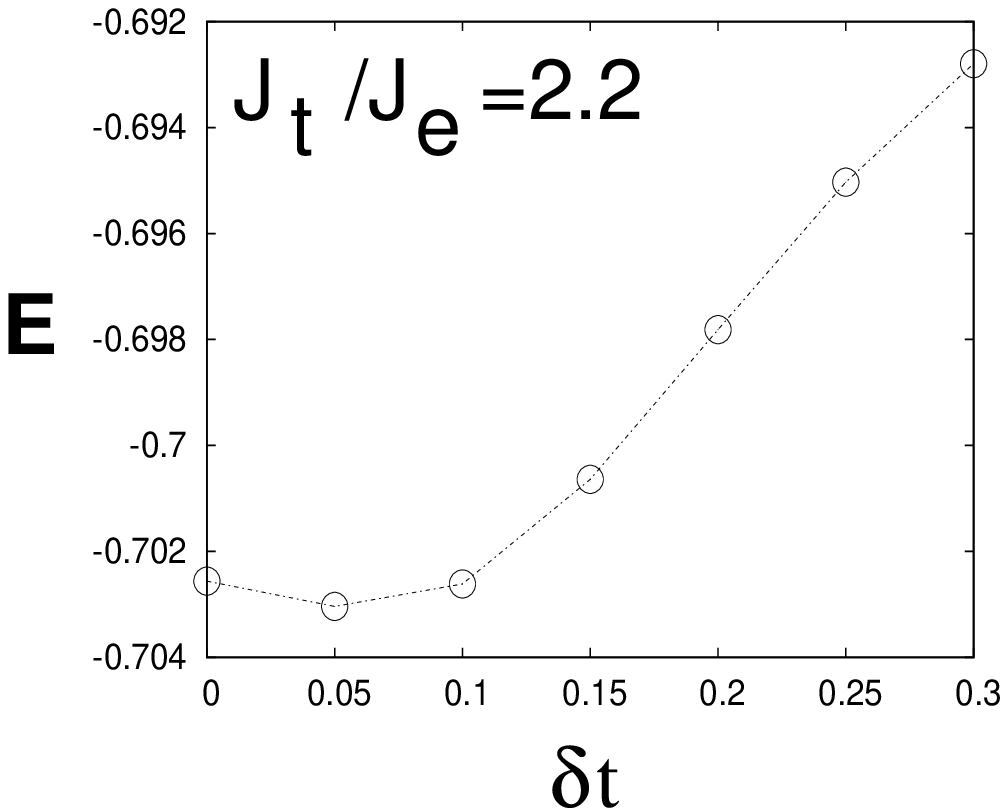}
\caption{Change in energy of the Gutzwiller projected SL$[0,0,\pi]$ spin liquid state upon including
a distortion $\delta t$ corresponding to increased fermion hopping amplitude on the dimerized bonds
of the $\sqrt{3}\times\sqrt{3}$ columnar dimer state shown in Fig.~\ref{fig:Jtdimerconfig}(a).
For $J_t/J_e=1.9,2.0$, the SL$[0,0,\pi]$ state
is found to be stable against this distortion, while it is seen to be unstable for $J_t/J_e=2.1,2.2$. (This
calculation was carried out on a system with $12\times 12$ unit cells, i.e., with $864$ spins.
The statistical errors on the computed energy are of the order of the symbol size.)
}
\label{fig:energy_dim}
\end{figure}

We next assess the stability of the SL$[0,0,\pi]$ spin liquid towards columnar
dimer order in the Guzwiller projected state.
In order to do this, we include a parameter $\delta t$ in the mean field Hamiltonian
which corresponds to strengthening the fermion hopping on those bonds which
dimerize in the `classical' columnar dimer state, shown in Fig.~\ref{fig:Jtdimerconfig}(a), and Gutzwiller
project the resulting state. Clearly, if $\delta t \! \gg \! 1$, the resulting wave function will
be precisely the `classical' columnar dimer pattern. The energy change of the weakly
distorted state
as a function of the distortion parameter $\delta t$ serves as a measure of the inverse
susceptibility of the SL$[0,0,\pi]$ state towards columnar dimer order. As seen from
Fig.~\ref{fig:energy_dim}, the SL$[0,0,\pi]$ state is stable, with a positive inverse
susceptibility, for $J_t/J_e \lesssim 2$, but is unstable, with a negative inverse susceptibility
for $J_t/J_e \gtrsim 2$. Further, the optimal $\delta t$ appears to increase continuously for
$J_t/J_e \gtrsim 2$. This suggests that the SL$[0,0,\pi]$ state possibly undergoes
a continuous transition into a state with $\sqrt{3}\times\sqrt{3}$ columnar dimer order at
$J_t/J_e \approx 2.0$. We discuss the phase transition more carefully in the next section.

\section{\label{sec:phasetransition} Phase transition between VBS phases}

The results in previous sections show that there is a phase transition between
the $J_{e}$-dimer VBS phase
and the columnar 18-site VBS phase with increasing $J_t/J_e$.
To describe the phase transition between these
two VBS phases, we construct a Landau-Ginzburg free energy
introducing a two-component vector ($\Phi_{1}$,$\Phi_{2}$), which transforms
as an $E_{3}$ irreducible representation of the enlarged point group $G_{P,\textbf{b}}$.
Since the VBS phases are time-reversal invariant, we can use two real numbers, $\Phi_{1}$ and $\Phi_{2}$.
The Landau-Ginzburg free energy can be written using all possible invariants made of
$\Phi_{1}$ and $\Phi_{2}$. In particular, it is important to note that
there is a third order invariant, which belongs to the $A_{1}$
irreducible representation of the following decomposition,
\begin{align}
E_{3}\otimes E_{3}\otimes E_{3}=A_{1}\oplus B_{1}\oplus 3 E_{3}.
\end{align}
In terms of $\Phi_{1}$ and $\Phi_{2}$, the third order invariant
is given by $\Phi_{2}(3\Phi_{1}^{2}-\Phi_{2}^{2})$.
Straightforward extension of the same group theoretical analysis to quartic order
shows that there is only one quartic invariant of $(\Phi_{1}^2+\Phi_{2}^2)^2$.
Collecting
all invariants up to quartic order, the Landau-Ginzburg free energy density
is written as
\begin{align}
f = \alpha (\Phi_{1}^2+\Phi_{2}^2)+\lambda \Phi_{2}(3\Phi_{1}^{2}-\Phi_{2}^{2})
+ u (\Phi_{1}^2+\Phi_{2}^2)^2.
\end{align}
For convenience we define a complex variable $\Phi$ as follows,
\begin{align}
\Phi  \equiv \Phi_{2}-i\Phi_{1} \equiv |\Phi|e^{i\theta}.
\end{align}
Then the free energy density is given by
\begin{align}
f = \alpha |\Phi|^2+ u |\Phi|^4-\lambda |\Phi|^{3}\cos(3\theta).
\end{align}
Given $u>0$ and $\lambda>0$,
the above mean field free energy predicts two different phases separated by a first order transition point
at $\alpha=\alpha_{c}=\lambda^{2}/(4\mu)$.
For $\alpha>\alpha_{c}$, we have a disordered phase with $|\Phi|=0$.
On the other hand, ordered phases with  $|\Phi|\neq0$ and $\theta=\frac{2\pi n}{3}$
(n=0,1,2) appear when $\alpha<\alpha_{c}$. The three ordered states
describe the three-fold degenerate VBS phases with $\sqrt{3}\times\sqrt{3}$ pattern.

Considering slow spatial variation of $\Phi$, the Euclidean Landau-Ginzburg
effective action is given by
\begin{align}
S = \int d^{3}x\Big\{|\partial_{\mu}\Phi|^2 +\alpha |\Phi|^2+ u |\Phi|^4-\lambda |\Phi|^{3}\cos(3\theta)\Big\}.
\end{align}
The above effective action is nothing but the action for the
three dimensional $Z_{3}$-clock model.
Previous Monte Carlo simulations suggest that the phase transition associated with the $Z_{3}$ symmetry breaking is weakly first order.\cite{Z3_1,Z3_2,Z3_3}
The wave function numerics in the previous section, however,
are not inconsistent with this scenario considering the small system size.
In addition, the nature of the transition may not have been completely settled.\cite{MYChoi}

\section{\label{sec:triplon} Triplon dispersion in the VBS states}
\begin{figure}[t]
\centering
\includegraphics[width=8.0 cm]{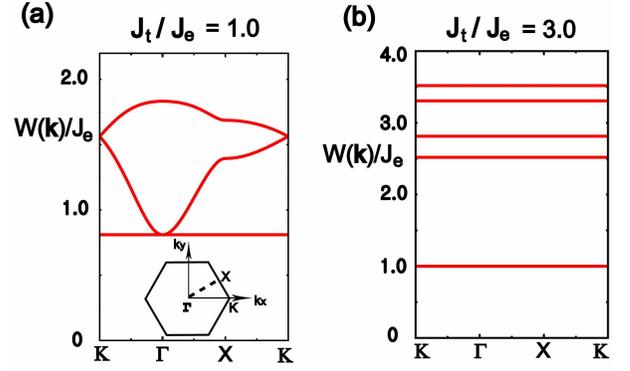}
\caption{Triplet dispersions of valence bond solid
ground states along high symmetry directions of the Brillouin zone.
The dispersions are obtained from
the self-consistent bond operator mean field theory.
(a) Triplet dispersion of the $J_{e}$-dimer VBS phase for $J_{t}/J_{e}=1$.
(b) Triplet dispersion of the columnar 18-site VBS phase for $J_{t}/J_{e}=3$.
}
\label{fig:tripletdispersion}
\end{figure}
In Fig.~\ref{fig:tripletdispersion},
we plot the triplet dispersions of the two VBS phases,
which are obtained from the bond operator mean field theory.
The triplet dispersion of the $J_{e}$-dimer VBS state
is shown in Fig.~\ref{fig:tripletdispersion} (a).
Since the unit cell of the $J_{e}$-dimer VBS state is composed
of three dimers, we have three triplet modes in the spectrum.
Here we neglect the fact that each triplet particle has three components (x, y, and z)
when we count the number of bands. Interestingly,
the lowest band is flat and touches another dispersive band
at the Brillouin zone center. The emergence of this band touching
has a topological origin.\cite{Bergman,Yang}
The flat band reflects
the existence of localized eigenstates.
In fact, there are two different types of localized eigenstates.
The first set is given by states which are confined
within dodecagons. Each dodecagon supports a single localized eigenstate.
In addition, there are ``non-contractible" loop states
constituting the second group of localized eigenstates.
In contrast to the states confined within dodecagons,
these loop states extend over the whole lattice system one-dimensionally.
The topological characteristics of these loop states can be easily understood
using the periodic boundary condition, under which a two-dimensional system
has a torus geometry. In this situation, there are two independent
loop states winding the torus once. Since these states cannot be shrunk to
points, they are ``non-contractible".
Counting the number of independent localized states carefully,
we see that the number of degenerate
localized eigenstates is larger than the number of dimers on the
lattice.\cite{Bergman,Yang}
It means that a single flat band is not enough to support all independent
localized eigenstates. Therefore additional degrees of freedom must
be provided by another band, leading to the band touching.

The triplet dispersion of the columnar 18-site VBS state is
displayed in Fig.~\ref{fig:tripletdispersion} (b).
We have nine triplet bands which are all flat within our mean field
approach.
The degeneracies of the flat bands are given
by 3, 1, 2, 2, and 1 counting from the bottom
to the top bands. In contrast to the case of the $J_e$-dimer VBS phase,
the flat structure emerges simply because the dimers
on expanded links are completely decoupled from
those on neighboring ``perfect" dodecagons.
The lowest flat band is triply degenerate, which comes
from the three dimers on expanded links.
The remaining six dimers of the unit cell lying on ``perfect" dodecagons constitute
the other six bands with higher energies.
Therefore
we expect that the nature of the valence bond solid ground states
can be demonstrated via neutron scattering experiments
measuring triplet dispersion spectra.

\section{\label{sec:discussion} Discussion}

In summary, we have shown that the ground state of the nearest neighbor Heisenberg model on the star
lattice undergoes a phase transition from  ``the $J_{e}$-dimer VBS phase" which respects all lattice symmetries
to ``the columnar 18-site VBS phase" which
exhibits $\sqrt{3}\times\sqrt{3}$ order with increasing $J_t/J_e$.
From the Landau-Ginzburg analysis, this
appears to be a conventional quantum phase transition
which is described as the thermal transition of the
$2+1$ dimensional $Z_{3}$-clock model.

If $S=1/2$
variants of the organic Iron-Acetate magnet can be synthesized, they would be
particularly good candidates to study the phase diagram discussed in this paper
since it
may be possible to pressure tune the ratio $J_t/J_e$ significantly in such systems.
Both VBS states
obtained here would exhibit a spin gap in uniform susceptibility measurements.
We expect a direct signature of the 18-site VBS order to appear in X-ray
scattering or neutron diffraction studies which would see a change in the crystal periodicity.
Ignoring coupling to phonons, the 18-site VBS state should exhibit a thermal transition in the
universality class of the $Z_{3}$-clock model in $D=2$ dimensions.
In addition, the two VBS phases exhibit quite distinct behaviors
in their triplet excitation spectra as discussed above which could be tested using
inelastic neutron scattering experiments.

\acknowledgments

We thank Daniel Podolsky, Tingpong Choy, Fa Wang, and Jason Alicea for helpful discussions.
This work was supported by the NSERC of Canada, the Canada Research Chair
Program and the Canadian Institute for Advanced Research.
AP acknowledges support from the Sloan Foundation, the Connaught Foundation
and an Ontario ERA.


\appendix

\section{Derivation of the low energy effective Hamiltonians}
Here we present the details of how we have derived the low energy effective Hamiltonians
of the spin liquid states discussed in Sec.~\ref{sec:properties}.

\subsection{\label{sec:effparallel} Effective Hamiltonian for SL$[\frac{\pi}{2} ,\frac{\pi}{2} , \pi]$}

For the gauge choice depicted in Fig.~\ref{fig:fluxconfig}(a), the mean field
Hamiltonian corresponding to the SL$[\frac{\pi}{2} ,\frac{\pi}{2} , \pi]$
can be written in momentum space as,

\begin{align}
H_{\text{MF}} &= - J_{t} \chi_{t} \sum_{\textbf{k}}\sum_{m, n}
f^{\dag}_{\textbf{k}, m} H(\textbf{k})_{m,n} f_{\textbf{k},n},
\end{align}
in which
\begin{displaymath}
H(\textbf{k})=
\left( \begin{array}{cccccc}
0 & i & i & \lambda z^{*}_{2} & 0 & 0 \\
-i & 0 & -i & 0 & \lambda & 0 \\
-i & i & 0 & 0 & 0 & \lambda z^{*}_{1} \\
\lambda z_{2} & 0 & 0 & 0 & -i & -i \\
0 & \lambda & 0 & i & 0 & -i \\
0 & 0 & \lambda z_{1} & i & i & 0
\end{array} \right),
\end{displaymath}
where $z_{1}$=$e^{i\textbf{k}\cdot\textbf{a}_{1}}$ and $z_{2}$=$e^{i\textbf{k}\cdot\textbf{a}_{2}}$.
$m$, $n$ are indices for the six sites inside a unit cell and $\lambda$=$J_{e}\chi_{e}$/$J_{t}\chi_{t}$.
The indices for the spin quantum number are dropped for simplicity.
Here we define the Fourier transformation via $f_{\textbf{R}, n}$ = $\frac{1}{\sqrt{N_{c}}}$$\sum_{\textbf{k}}$
$e^{i \textbf{k}\cdot \textbf{R}}$$f_{\textbf{k},n}$.

As described in Fig.~\ref{fig:fluxband}(b), the conduction (valence) band shows
the dispersion minimum (maximum) at the momentum $\pm \textbf{Q}$. The energy eigenvalues
of $H(\textbf{k})$ at $\textbf{k}$=$\pm \textbf{Q}$ are given by

\begin{align}
E^{\pm}_{1} &= \sqrt{3/2+\lambda^{2}-1/2 \sqrt{9+12 \lambda^{2}}},\nonumber\\
E^{\pm}_{2} &= -E^{\pm}_{1},
\end{align}
where $\pm$ indicates the two momentum position $\pm \textbf{Q}$ and 1 and 2
represent the conduction (1) and valence (2) bands, respectively.
To make the analytic treatment of the problem possible we focus on
the small $\lambda$ limit. Note that the overall spinon band structure
does not change upon varying $\lambda$. Expanding the energy eigenvalues in powers of
$\lambda$, we get

\begin{align}
E^{\pm}_{1} &= - E^{\pm}_{2} \approx
\frac{1}{\sqrt{3}}\lambda^{2} + O(\lambda^{4}).\nonumber
\end{align}

The corresponding eigenvectors are

\begin{align}\label{eq:eigenvector1}
(\nu^{+}_{1})^{T} &=\frac{e^{-i \pi/6}}{\sqrt{3(1+b^{2})}}\{-1-\frac{1}{3}\lambda^{2},
-1-\frac{1}{3}\lambda^{2},1+\frac{1}{3}\lambda^{2},\nonumber\\
&\qquad\qquad\qquad\qquad
-\frac{1}{\sqrt{3}}\lambda e^{i \pi/3},
-\frac{1}{\sqrt{3}}\lambda,\frac{1}{\sqrt{3}}\lambda e^{i 2\pi/3}\},\nonumber\\
(\nu^{+}_{2})^{T} &=\frac{e^{i \pi/6}}{\sqrt{3(1+b^{2})}}\{-\frac{1}{\sqrt{3}}\lambda e^{-i \pi/3},
\frac{1}{\sqrt{3}}\lambda,-\frac{1}{\sqrt{3}}\lambda e^{-i 2\pi/3},\nonumber\\
&\qquad\qquad\qquad\qquad
1+\frac{1}{3}\lambda^{2},
-1-\frac{1}{3}\lambda^{2},1+\frac{1}{3}\lambda^{2}\},\nonumber\\
(\nu^{-}_{1})^{T} &=\frac{e^{-i 2\pi/3}}{\sqrt{3(1+b^{2})}}\{\frac{1}{\sqrt{3}}\lambda e^{i \pi/3},
-\frac{1}{\sqrt{3}}\lambda,\frac{1}{\sqrt{3}}\lambda e^{i 2\pi/3},\nonumber\\
&\qquad\qquad\qquad\qquad
1+\frac{1}{3}\lambda^{2},
-1-\frac{1}{3}\lambda^{2},1+\frac{1}{3}\lambda^{2}\},\nonumber\\
(\nu^{-}_{2})^{T} &=\frac{e^{i 2\pi/3}}{\sqrt{3(1+b^{2})}}\{-1-\frac{1}{3}\lambda^{2},
-1-\frac{1}{3}\lambda^{2},1+\frac{1}{3}\lambda^{2},\nonumber\\
&\qquad\qquad\qquad\qquad
\frac{1}{\sqrt{3}}\lambda e^{-i \pi/3},
\frac{1}{\sqrt{3}}\lambda,-\frac{1}{\sqrt{3}}\lambda e^{-i 2\pi/3}\},
\end{align}
where the superscript $T$ means taking transposition. The above
eigenvalues and eigenvectors satisfy $H(\pm \textbf{Q})$$\nu^{\pm}_{\alpha}$
=$E^{\pm}_{\alpha}$$\nu^{\pm}_{\alpha}$ ($\alpha$=1 and 2) correctly
up to the third order in $\lambda$.

Now we want to construct the effective Hamiltonian describing the
states which have small momentum deviation from $\pm \textbf{Q}$, i.e.,
states with $\textbf{k}$=$\pm \textbf{Q}$+$\textbf{q}$.
We first define $\Delta H(\textbf{q})$ $\equiv$ $H(\pm \textbf{Q} + \textbf{q})$-$H(\pm \textbf{Q})$.
Keeping terms which are first order in $\textbf{q}$ and projecting them into the low energy space
spanned by the eigenvectors $\nu^{\pm}_{\alpha}$ ($\alpha$ = 1, 2), we obtain the
following effective Hamiltonian,

\begin{align}
H^{\pm}(\textbf{q}) & \equiv  H(\pm \textbf{Q}+\textbf{q})
= - \frac{\lambda}{\sqrt{3}}(q_{x}\tau_{x} + q_{y}\tau_{y})+\frac{\lambda^{2}}{\sqrt{3}}\tau_{z}.\nonumber
\end{align}

Finally, defining the continuum fermion fields using the spinon variables as

\begin{align}\label{eq:continuumfield}
(\psi_{\pm}(\textbf{q}))^{T} &\sim \{\sum_{n=1}^{6}(\nu^{\pm}_{1})^{*}_{n} f_{\pm \textbf{Q} + \textbf{q},n},
\sum_{n=1}^{6}(\nu^{\pm}_{2})^{*}_{n} f_{\pm \textbf{Q} + \textbf{q},n}\},\nonumber\\
(\Psi(\textbf{q}))^{T} &\equiv ((\psi_{+}(\textbf{q}))^{T},(\psi_{-}(\textbf{q}))^{T}),
\end{align}
we arrive at the low energy effective Hamiltonian written in Eq.~(\ref{eq:EqDirac1})
which is nothing but the Hamiltonian for the massive Dirac particles.

\subsection{\label{sec:effstaggered} Effective Hamiltonian for SL$[-\frac{\pi}{2} ,\frac{\pi}{2} , 0]$}
The low energy Hamiltonian corresponding to the SL$[-\frac{\pi}{2} ,\frac{\pi}{2} , 0]$
can be obtained following the similar steps used to construct the massive Dirac
Hamiltonian of the  SL$[\frac{\pi}{2} ,\frac{\pi}{2} , \pi]$.
Adopting the flux configuration depicted in Fig.~\ref{fig:fluxconfig}(b),
we have an electron pocket centered at the momentum $\textbf{K}$ and an hole pocket
centered at the momentum -$\textbf{K}$. At each momentum $\pm \textbf{K}$ there is a linear
band touching as is shown in Fig.~\ref{fig:fluxconfig}(b). The energy eigenvalues
of the degenerate states at $\textbf{k}$=$\pm \textbf{K}$ are given by

\begin{align}
E^{+} &= \frac{\sqrt{3+\lambda^{2}}-\sqrt{3}}{2} \approx
\frac{1}{\sqrt{3}}\lambda^{2} + O(\lambda^{4}),\nonumber\\
E^{-} &= -E^{+} \approx
-\frac{1}{\sqrt{3}}\lambda^{2} + O(\lambda^{4}),\nonumber
\end{align}
where $E^{+}$ ($E^{-}$) represents the degenerate energy eigenvalue at the momentum
$\textbf{K}$ (-$\textbf{K}$). We choose the corresponding eigenvectors in the
following way,

\begin{align}\label{eq:eigenvector2}
(\nu^{+}_{1})^{T} &=\frac{e^{i \pi/3}}{\sqrt{3(1+b^{2})}}\{-\frac{1}{\sqrt{3}}\lambda e^{-i 2\pi/3},
-\frac{1}{\sqrt{3}}\lambda,\frac{1}{\sqrt{3}}\lambda e^{i 2\pi/3},\nonumber\\
&\qquad\qquad\qquad\qquad
-1-\frac{1}{3}\lambda^{2},
-1-\frac{1}{3}\lambda^{2},1+\frac{1}{3}\lambda^{2}\},\nonumber\\
(\nu^{+}_{2})^{T} &=\frac{e^{-i \pi/3}}{\sqrt{3(1+b^{2})}}\{-1-\frac{1}{3}\lambda^{2},
-1-\frac{1}{3}\lambda^{2},1+\frac{1}{3}\lambda^{2},\nonumber\\
&\qquad\qquad\qquad\qquad
-\frac{1}{\sqrt{3}}\lambda e^{i 2\pi/3},
-\frac{1}{\sqrt{3}}\lambda,\frac{1}{\sqrt{3}}\lambda e^{-i 2\pi/3}\},\nonumber\\
(\nu^{-}_{1})^{T} &=\frac{e^{-i \pi/6}}{\sqrt{3(1+b^{2})}}\{-1-\frac{1}{3}\lambda^{2},
-1-\frac{1}{3}\lambda^{2},1+\frac{1}{3}\lambda^{2},\nonumber\\
&\qquad\qquad\qquad\qquad
\frac{1}{\sqrt{3}}\lambda e^{-i 2\pi/3},
\frac{1}{\sqrt{3}}\lambda,-\frac{1}{\sqrt{3}}\lambda e^{i 2\pi/3}\},\nonumber\\
(\nu^{-}_{2})^{T} &=\frac{e^{i \pi/6}}{\sqrt{3(1+b^{2})}}\{\frac{1}{\sqrt{3}}\lambda e^{i 2\pi/3},
\frac{1}{\sqrt{3}}\lambda,-\frac{1}{\sqrt{3}}\lambda e^{-i 2\pi/3},\nonumber\\
&\qquad\qquad\qquad\qquad
-1-\frac{1}{3}\lambda^{2},
-1-\frac{1}{3}\lambda^{2},1+\frac{1}{3}\lambda^{2}\},
\end{align}
satisfying $H(\pm \textbf{K})$$\nu^{\pm}_{\alpha}$
=$E^{\pm}$$\nu^{\pm}_{\alpha}$ ($\alpha$=1 and 2) correctly
up to the third order in $\lambda$. The first order perturbation
theory combined with the projection into the low energy space
spanned by $\nu^{\pm}_{\alpha}$ ($\alpha$=1, 2) leads to the
following Hamiltonian,

\begin{align}
H^{\pm}(\textbf{q}) & \equiv  H(\pm \textbf{K}+\textbf{q})
= - \frac{\lambda}{\sqrt{3}}(q_{x}\tau_{x} + q_{y}\tau_{y})+\frac{\lambda^{2}}{\sqrt{3}}\mu_{z}.\nonumber
\end{align}

Using the continuum fermion field defined in Eq.~(\ref{eq:continuumfield})
we obtain the low energy effective Hamilonian displayed in Eq.~(\ref{eq:effectiveH2})

\subsection{\label{sec:eff000}Effective Hamiltonian for SL$[0 ,0 ,0]$}

The mean field band structure of the SL$[0 ,0 ,0]$ has a flat band lying
at the fermi energy, which is touching a dispersive band at the zone
center, i.e., at the momentum $\mathbf{\Gamma}$ = (0,0). To describe the low energy states near the
$\mathbf{\Gamma}$ point, we use the degenerate perturbation theory again. However,
since the two bands touch quadratically, we keep the perturbation expansion
up to the quadratic order in momentum.
For the SL$[0 ,0 ,0]$, the mean field Hamiltonian is given by

\begin{align}
H_{\text{MF}} &= - J_{t} \chi_{t} \sum_{\textbf{k}}\sum_{m, n}
f^{\dag}_{\textbf{k}, m} H(\textbf{k})_{m,n} f_{\textbf{k},n},
\end{align}
in which
\begin{displaymath}
H(\textbf{k})=
\left( \begin{array}{cccccc}
0 & 1 & 1 & \lambda z^{*}_{2} & 0 & 0 \\
1 & 0 & 1 & 0 & \lambda & 0 \\
1 & 1 & 0 & 0 & 0 & \lambda z^{*}_{1} \\
\lambda z_{2} & 0 & 0 & 0 & 1 & 1 \\
0 & \lambda & 0 & 1 & 0 & 1 \\
0 & 0 & \lambda z_{1} & 1 & 1 & 0
\end{array} \right).
\end{displaymath}

We divide $H(\textbf{k})$ into two pieces such that $H(\textbf{k})$ = $H_{0}$ + $V$
in which $H_{0}$ $\equiv$ $H(\textbf{k}=0)$. Diagonalization of $H_{0}$ gives
the eigenvalues $E_{0}$=$\{$-1-$\lambda$,-1-$\lambda$,-1+$\lambda$,-1+$\lambda$,2-$\lambda$,2+$\lambda$$\}$.
Note that there are two pairs of doubly degenerate eigenvalues.
Here we focus on one of the degenerate eigenvalues $w_{0}$=-1+$\lambda$
which is lying at the fermi level. We choose the following two degenerate
eigenvectors corresponding to $w_{0}$,

\begin{align}
(\nu_{1})^{T} &= \frac{1}{\sqrt{12}}\{2, -1, -1, 2, -1, -1\},\nonumber\\
(\nu_{2})^{T} &= \frac{1}{\sqrt{4}}\{0, 1, -1, 0, 1, -1\},\nonumber
\end{align}

Now we introduce the projection operator $\hat{P}_{0}$ ($\hat{P}_{1}$)
which projects states into (out of) the low energy space
spanned by $\nu_{1}$ and $\nu_{2}$. That is, $\hat{P}_{0}$=$\nu_{1}\cdot\nu^{\dag}_{1}$ + $\nu_{2}\cdot\nu^{\dag}_{2}$
and $\hat{P}_{1}$ = $\hat{I}$ - $\hat{P}_{0}$. We also define $V_{0}$ $\equiv$ $\hat{P}_{0}V\hat{P}_{0}$
and $V_{1}$ $\equiv$ $V$ -$V_{0}$. Then $H(\textbf{k})$ = $H_{0}$ + $V_{0}$ + $V_{1}$.
The projected Hamiltonian is given by\cite{Sakurai}

\begin{align}
H_{P_{0}}(\textbf{k})&\equiv \hat{P}_{0}H(\textbf{k})\hat{P}_{0}\nonumber\\
&\approx  \hat{P}_{0}[H_{0} + V_{0}]\hat{P}_{0} +
\hat{P}_{0}V_{1}\hat{P}_{1}\frac{1}{w_{0}-H_{0}}\hat{P}_{1}V_{1}\hat{P}_{0},
\end{align}
which is valid up to the quadratic order in $\textbf{k}$. The resulting Hamiltonian
can be written as,

\begin{displaymath}
H_{P_{0}}(\textbf{k})=
\frac{\lambda}{2\lambda-3}
\left( \begin{array}{cc}
k^{2}_{y} & -k_{x}k_{y} \\
-k_{x}k_{y} & k^{2}_{x}
\end{array} \right),
\end{displaymath}

We define the continuum fields as

\begin{align}
(\psi(\textbf{k}))^{T} &\sim \{\sum_{n=1}^{6}(\nu^{\pm}_{1})^{*}_{n} f_{\textbf{k},n},
\sum_{n=1}^{6}(\nu^{\pm}_{2})^{*}_{n} f_{\textbf{k},n}\}.
\end{align}

Finally, combining the above results we obtain the following low energy
effective Hamiltonian,

\begin{align}
H_{\text{eff}} = &\frac{1}{m_{\text{eff}}}\int \frac{d^{2}\textbf{k}}{(2\pi)^2} \psi^{\dag}(\textbf{k})
h_{\text{eff}}(\textbf{k})\psi(\textbf{k}),\nonumber
\end{align}
in which
\begin{align}
h_{\text{eff}}(\textbf{k}) = &
(k^{2}_{x}+k^{2}_{y})\tau_{0}-(k^{2}_{x}-k^{2}_{y})\tau_{z}-2k_{x}k_{y}\tau_{x}.\nonumber
\end{align}

\section{\label{sec:fieldsymmetry} Symmetry and continuum field of the SL$[-\frac{\pi}{2} ,\frac{\pi}{2} , 0]$ state}

In this section we show how the continuum fields of the SL$[-\frac{\pi}{2} ,\frac{\pi}{2} , 0]$
state transform under the microscopic
symmetries of the lattice.
Here we follow the procedures suggested by Hermele et al. in Ref.~\onlinecite{Hermele}.
To obtain the necessary information  we consider
a finite system with periodic boundary conditions in both $\textbf{a}_{1}$ and $\textbf{a}_{2}$ directions.
In particular, to determine how the wave functions at the momentum $\pm \textbf{K}$ (we call it as the nodal wave functions)
transform under the space group symmetries of the ansatz, we consider the 3$\times$3 lattice system, that is,
the system is periodic under the translation by 3$\textbf{a}_{1}$ and 3$\textbf{a}_{2}$.
Since a unit cell (indexed by a vector $\textbf{R}$) contains six sites labeled by n, the finite system consists of
54 sites. The nine points within the Brillouin zone of the finite system
contain the the nodal points $\pm \textbf{K}$ and respect all the point group
symmetries of the ansatz.

Using the eigenvectors $\nu^{a}_{\alpha}$ (a = +, - and $\alpha$ = 1, 2)
in Eq.~(\ref{eq:eigenvector2}), the nodal wave function is given by

\begin{align}
\Phi_{a,\alpha}(\textbf{R},n) &= \frac{e^{\textbf{iaK}\cdot \textbf{R}}(\nu^{a}_{\alpha})_{n}}{3}.\nonumber
\end{align}

Then the continuum field is written as

\begin{align}
\Psi_{a,\alpha}(\textbf{q}=0) &= \sum_{\textbf{R},n} \Phi^{*}_{a,\alpha}(\textbf{R},n)f_{\textbf{R},n}.\nonumber
\end{align}

Now we consider a symmetry operation $S$ under which the spinons transform as

\begin{align}
S : & f_{\textbf{i}, \sigma} \rightarrow G_{S}(\textbf{i})f_{S(\textbf{i}), \sigma}. \nonumber
\end{align}

In particular we consider the following symmetry operations,

(i) The $\textbf{a}_{1}$ translation ($T_{1}$);
\begin{align}
T_{1} : & f_{\textbf{R}, n, \sigma} \rightarrow G_{T_{1}}(\textbf{R}, n)f_{\textbf{R}+\textbf{a}_{1}, n, \sigma}, \nonumber
\end{align}

(ii) The $\textbf{a}_{2}$ translation ($T_{2}$);
\begin{align}
T_{2} : & f_{\textbf{R}, n, \sigma} \rightarrow G_{T_{2}}(\textbf{R}, n)f_{\textbf{R}+\textbf{a}_{2}, n, \sigma}, \nonumber
\end{align}

(iii) The $\frac{2\pi}{3}$ rotation ($C_{2\pi/3}$);
\begin{align}
C_{2\pi/3} : & f_{\textbf{R}, n, \sigma} \rightarrow G_{C_{2\pi/3}}(\textbf{R}, n)f_{C_{2\pi/3}(\textbf{R}, n), \sigma}, \nonumber
\end{align}

(iv) The y reflection ($R_{y}$);
\begin{align}
R_{y} : & f_{(x,y), n, \sigma} \rightarrow G_{R_{y}}(\textbf{R}, n)f_{(x,-y), m, \sigma}, \nonumber
\end{align}
where $\textbf{R}$=(x,y) and (n,m)$\in$$\{(1,4),(2,6),(3,5),(4,1),(5,3),(6,2)\}$.

(v) Time-reversal and inversion ($T \cdot I$);
\begin{align}
T \cdot I : & f_{\textbf{R}, n, \sigma} \rightarrow (i \sigma_{2})_{\sigma,\sigma'}f_{-\textbf{R}, (n+3), \sigma'}, \nonumber
\end{align}
in which (n+3) represents the remainder when it is divided by six.

(vi) Charge conjugation ($C^{*}$);
\begin{align}
C_{*} : & f_{\textbf{R}, n, \sigma} \rightarrow \epsilon_{n}f^{\dag}_{\textbf{R}, n, \sigma}, \nonumber
\end{align}
in which $\epsilon_{i}$ = 1 for i = 1, 2, 3 and -1 otherwise.

For each of the symmetry operation $S$, the matrix representation of the symmetry $S$
is defined as

\begin{align}
(S)_{S(\textbf{i}),\textbf{i}}& \equiv G_{S}(\textbf{i}). \nonumber
\end{align}

The action of the symmetry operation $S$ on the nodal wave function is given by
$S\Phi_{a}$=$c_{ab}\Phi_{b}$. Now a and b denote the nodal and the two-component
Dirac indices collectively. Finally, the transformation of the nodal wave function
reflects the action of $S$ on the continuum field such as

\begin{align}
S : \Psi_{a} &\rightarrow c^{*}_{ab}\Psi_{b}. \nonumber
\end{align}

The transformation rule of the continuum field under all the
above symmetry operations is summarized in Eq.~(\ref{eq:symmetryofcontinuum}).

\section{\label{sec:matrices} Expressions of the matrices $A_{S}(\Pi_{i})$ and $B_{S}(\Omega_{i})$}

Here we present the expressions of the matrices $A_{S}(\Pi_{i})$
and $B_{S}(\Omega_{i}) $ ($S$=$T_{a_1}$, $T_{a_2}$, $R_{y}$, and $C_{\frac{\pi}{3}}$),
which are defined in Sec.~\ref{sec:psg}. At first $A_{S}(\Pi_{1})$ are given by

\begin{displaymath}
A_{T_{a_{1}}}(\Pi_1) :
\left( \begin{array}{cccccc}
e^{i\frac{2\pi}{3}} & 0 & 0 & 0 & 0 & 0 \\
0 &e^{-i\frac{\pi}{3}} & 0 & 0 & 0 & 0  \\
0 & 0 &e^{-i\frac{\pi}{3}} & 0 & 0 & 0 \\
0 & 0 & 0 &e^{-i\frac{2\pi}{3}} & 0 & 0 \\
0 & 0 & 0 & 0 &e^{i\frac{\pi}{3}} & 0 \\
0 & 0 & 0 & 0 & 0 &e^{i\frac{\pi}{3}}
\end{array} \right),
\end{displaymath}

\begin{displaymath}
A_{T_{a_{2}}}(\Pi_1) :
\left( \begin{array}{cccccc}
e^{i\frac{\pi}{3}} & 0 & 0 & 0 & 0 & 0 \\
0 &e^{i\frac{\pi}{3}} & 0 & 0 & 0 & 0  \\
0 & 0 &e^{-i\frac{2\pi}{3}} & 0 & 0 & 0 \\
0 & 0 & 0 &e^{-i\frac{\pi}{3}} & 0 & 0 \\
0 & 0 & 0 & 0 &e^{-i\frac{\pi}{3}} & 0 \\
0 & 0 & 0 & 0 & 0 &e^{i\frac{2\pi}{3}}
\end{array} \right),
\end{displaymath}

\begin{displaymath}
A_{R_{y}}(\Pi_1) :
\left( \begin{array}{cccccc}
1 & 0 & 0 & 0 & 0 & 0 \\
0 & 0 & 1 & 0 & 0 & 0  \\
0 & 1 & 0 & 0 & 0 & 0 \\
0 & 0 & 0 & 1 & 0 & 0 \\
0 & 0 & 0 & 0 & 0 & -1 \\
0 & 0 & 0 & 0 & -1 & 0
\end{array} \right),
\end{displaymath}

\begin{displaymath}
A_{C_{\frac{\pi}{3}}}(\Pi_1) :
\left( \begin{array}{cccccc}
0 & 0 & 0 & 0 & 0 & 1 \\
0 & 0 & 0 & 1 & 0 & 0  \\
0 & 0 & 0 & 0 & i & 0 \\
0 & 0 & 1 & 0 & 0 & 0 \\
-1 & 0 & 0 & 0 & 0 & 0 \\
0 & i & 0 & 0 & 0 & 0
\end{array} \right).
\end{displaymath}

Similarly for $A_{S}(\Pi_{2})$,
\begin{displaymath}
A_{T_{a_{1}}}(\Pi_2) :
\left( \begin{array}{cccccc}
-1 & 0 & 0 & 0 & 0 & 0 \\
0 & 1 & 0 & 0 & 0 & 0  \\
0 & 0 & -1 & 0 & 0 & 0 \\
0 & 0 & 0 & -1 & 0 & 0 \\
0 & 0 & 0 & 0 & 1 & 0 \\
0 & 0 & 0 & 0 & 0 & -1
\end{array} \right),
\end{displaymath}

\begin{displaymath}
A_{T_{a_{2}}}(\Pi_2) :
\left( \begin{array}{cccccc}
-1 & 0 & 0 & 0 & 0 & 0 \\
0 & -1 & 0 & 0 & 0 & 0  \\
0 & 0 & 1 & 0 & 0 & 0 \\
0 & 0 & 0 & -1 & 0 & 0 \\
0 & 0 & 0 & 0 & -1 & 0 \\
0 & 0 & 0 & 0 & 0 & 1
\end{array} \right),
\end{displaymath}

\begin{displaymath}
A_{R_{y}}(\Pi_2) :
\left( \begin{array}{cccccc}
0 & 0 & 1 & 0 & 0 & 0 \\
0 & -1 & 0 & 0 & 0 & 0  \\
1 & 0 & 0 & 0 & 0 & 0 \\
0 & 0 & 0 & 0 & 0 & 1 \\
0 & 0 & 0 & 0 & -1 & 0 \\
0 & 0 & 0 & 1 & 0 & 0
\end{array} \right),
\end{displaymath}

\begin{displaymath}
A_{C_{\frac{\pi}{3}}}(\Pi_2) :
\left( \begin{array}{cccccc}
0 & 0 & 0 & 0 & i & 0 \\
0 & 0 & 0 & 0 & 0 & i  \\
0 & 0 & 0 & 1 & 0 & 0 \\
0 & i & 0 & 0 & 0 & 0 \\
0 & 0 & i & 0 & 0 & 0 \\
1 & 0 & 0 & 0 & 0 & 0
\end{array} \right).
\end{displaymath}

In the case of $B_{S}(\Omega_1)$,

\begin{displaymath}
B_{T_{a_{1}}}(\Omega_1) :
\left( \begin{array}{cc}
e^{i\frac{2\pi}{3}} & 0  \\
0 &e^{-i\frac{2\pi}{3}}
\end{array} \right),\quad
B_{T_{a_{2}}}(\Omega_1) :
\left( \begin{array}{cc}
e^{i\frac{-2\pi}{3}} & 0  \\
0 &e^{i\frac{2\pi}{3}}
\end{array} \right),
\end{displaymath}

\begin{displaymath}
B_{R_{y}}(\Omega_1) :
\left( \begin{array}{cc}
-1 & 0  \\
0 & -1
\end{array} \right),\quad
B_{C_{\frac{\pi}{3}}}(\Omega_1) :
\left( \begin{array}{cc}
0 & -i  \\
i & 0
\end{array} \right).
\end{displaymath}

Similarly for $B_{S}(\Omega_2)$,
\begin{displaymath}
B_{T_{a_{1}}}(\Omega_2) :
\left( \begin{array}{cc}
1 & 0  \\
0 & 1
\end{array} \right),\quad
B_{T_{a_{2}}}(\Omega_2) :
\left( \begin{array}{cc}
1 & 0  \\
0 & 1
\end{array} \right),
\end{displaymath}

\begin{displaymath}
B_{R_{y}}(\Omega_2) :
\left( \begin{array}{cc}
1 & 0  \\
0 & 1
\end{array} \right),\quad
B_{C_{\frac{\pi}{3}}}(\Omega_2) :
\left( \begin{array}{cc}
0 & 1  \\
1 & 0
\end{array} \right).
\end{displaymath}



\end{document}